\documentclass[a4paper,twoside,runningheads]{llncs}
\usepackage[T1]{fontenc}
\usepackage{times}
\usepackage[hypertexnames=false]{hyperref}
\usepackage{amsmath}
\usepackage{amssymb}
\usepackage{amsthm}
\usepackage[dvipsnames]{xcolor}
\usepackage{pdfpages}

\usepackage{stmaryrd}
\usepackage{laws}

\newif\ifarxiv
\arxivtrue
\newif\ifreview
\reviewtrue

\allowdisplaybreaks
\ifarxiv
\reviewfalse
\pdfoutput=1 
\else
\allowdisplaybreaks
\fi

\usepackage{imakeidx}
\makeindex

\newcommand{\draftonly}[1]{#1}

\newcommand\numberthis{\addtocounter{equation}{1}\tag{\theequation}}

\mathchardef\mhyphen="2D
\newenvironment{RelatedWork}{\par\color{blue}\textbf{Related work.}}{}

\newcommand{\C}{C}
\newcommand{\refsto}{\mathrel{\succeq}}

\newcommand{\nondet}{\mathbin{\vee}}
\newcommand{\Nondet}{\mathop{\textstyle\bigvee}}
\newcommand{\meet}{\mathbin{\wedge}}
\newcommand{\Meet}{\mathop{\textstyle\bigwedge}}
\newcommand{\together}{\mathbin{\Cap}}
\newcommand{\Seq}{\mathbin{;}}
\newcommand{\sync}{\mathbin{\otimes}}
\newcommand{\Fin}[1]{#1^{\star}}
\newcommand{\Inf}[1]{#1^{\infty}}
\newcommand{\Om}[1]{#1^{\omega}}

\newcommand{\Ata}{\mathsf{a}}
\newcommand{\Atb}{\mathsf{b}}
\newcommand{\Patx}{\mathsf{x}}

\newcommand{\kw}[1]{\mathsf{#1}}
\newcommand{\Abort}{\top}
\newcommand{\Assert}[1]{\mathop{\kw{assert}} #1}
\newcommand{\Chaos}{\kw{chaos}}
\newcommand{\Skip}{\kw{skip}}
\newcommand{\Guarantee}{\mathop{\kw{guar_{\pstepd}}}}
\newcommand{\GuarAll}{\mathop{\kw{guar_{\stepd}}}}
\newcommand{\guar}[1]{\Guarantee #1}
\newcommand{\Evolve}{\mathop{\kw{evolve}}}
\newcommand{\evolve}[1]{\Evolve #1}
\newcommand{\guarall}[1]{\GuarAll #1}
\newcommand{\Magic}{\bot}
\newcommand{\Nil}{\boldsymbol{\tau}}
\newcommand{\pre}[1]{\mathop{\kw{pre}}#1}
\newcommand{\notimmedabort}[1]{\mathop{not\_immed\_abort}#1}
\newcommand{\Pre}[1]{\{\hspace{-2pt}|#1|\hspace{-2pt}\}}
\newcommand{\Post}[1]{\Spec{}{}{#1}}
\newcommand{\Approx}{\mathbin{\downharpoonleft}}

\makeatletter
\def\Set{\@ifnextchar*{\@Set}{\@@Set}}
\def\@Set*#1{{\color{blue}\left\llcorner\begin{array}{l}#1\end{array}\right\lrcorner}}
\def\@@Set#1{{\color{blue}\llcorner#1\lrcorner}}
\def\Rel{\@ifnextchar*{\@Rel}{\@@Rel}}
\def\@Rel*#1{{\color{blue}\left\ulcorner\begin{array}[t]{l}#1\end{array}\right\urcorner}}
\def\@@Rel#1{{\color{blue}\ulcorner#1\urcorner}} 
\def\RelA{\@ifnextchar*{\@RelA}{\@@RelA}}
\def\@RelA*#1{{\color{purple}\left\lceil\BB\left\lceil\begin{array}{l}#1\end{array}\right\rceil\BB\right\rceil}}
\def\@@RelA#1{{\color{purple}\lceil\BB\lceil#1\rceil\BB\rceil}}
\def\Ratomicrel{\@ifnextchar*{\@Ratomicrel}{\@@Ratomicrel}}
\def\@Ratomicrel*#1{\left\langle\Rel*{#1}\right\rangle}
\def\@@Ratomicrel#1{\atomicrel{\Rel{#1}}}
\def\Spre{\@ifnextchar*{\@Spre}{\@@Spre}}
\def\@Spre*#1{\left\{\Set*{#1}\right\}}
\def\@@Spre#1{\{\Set{#1}\}}
\def\Spec{\@ifnextchar*{\@Spec}{\@@Spec}}
\def\@Spec*#1#2#3{\ifx\@empty#1\else#1:\fi
   \llparenthesis{#2}\ifx\@empty#2\else,~\fi#3\rrparenthesis}
\def\@@Spec#1#2#3{\ifx\@empty#1\else
   \begin{array}{@{}l@{}}#1\end{array}:\fi%
   \left(\hspace*{-2pt}\left|{\begin{array}{@{}l@{}}#2\end{array}}\ifx\@empty#2\else~,~~\fi
   \begin{array}{@{}l@{}}#3\end{array}\right|\hspace*{-2pt}\right)}
\makeatother

\newcommand{\Rely}{\mathop{\kw{rely}}}
\newcommand{\rely}[1]{\Rely #1}
\newcommand{\Rrely}[1]{\rely{\Rel{#1}}}
\newcommand{\Rguar}[1]{\guar{\Rel{#1}}}

\newcommand{\Rpost}[1]{\Post{\Rel{#1}}}

\newcommand{\cgd}[1]{\mathop{\tau} #1}
\newcommand{\pstepd}{\pi}
\newcommand{\estepd}{\epsilon}
\newcommand{\stepd}{\alpha}
\newcommand{\cstepd}{\boldsymbol{\stepd}}

\newcommand{\cpstep}[1]{\mathop{\pstepd} #1}
\newcommand{\cestepd}{\boldsymbol{\estepd}}
\newcommand{\cestep}[1]{\mathop{\estepd} #1}

\newcommand{\Atomic}{Atom}
\newcommand{\Command}{\mathcal{C}}

\newcommand{\defs}{\mathrel{\widehat=}}
\newcommand{\inter}{\mathbin{\cap}}
\newcommand{\nat}{\mathbb{N}}
\newcommand{\spot}{\mathrel{.}}
\newcommand{\union}{\mathbin{\cup}}
\newcommand{\universalset}{\mathsf{univ}}
\newcommand{\universalrel}{\mathsf{univ}}
\newcommand{\upto}{\mathbin{..}}

\makeatletter
\newcounter{colwidth}
\newenvironment{eqncolumns}{\@ifnextchar[{\@eqncolumns}{\@@eqncolumns}}{\end{eqnarray}\end{minipage}\vspace{1ex}\par}
\def\@eqncolumns[#1]{\setcounter{colwidth}{100-#1}\mbox{}\vspace{-4ex}\\\begin{minipage}[t]{0.#1\textwidth}\begin{eqnarray}}
\def\@@eqncolumns{\setcounter{colwidth}{5}\mbox{}\vspace{-4ex}\\\begin{minipage}[t]{0.5\textwidth}\begin{eqnarray}}
\newcommand{\secondcolumn}{\end{eqnarray}\end{minipage}\begin{minipage}[t]{0.\thecolwidth\textwidth}\begin{eqnarray}}
\makeatother

\makeatletter
\newcommand{\ChainRel}[1]{\crcr \noalign{\penalty\interdisplaylinepenalty}
  \hspace*{-1em}#1~ &
  \@ifnextchar*{\@ChainRelCommment}{}}
\newcommand{\Why}[1]{\mbox{{\color{blue}\hspace*{1em}#1}}}
\def\@ChainRelCommment*[#1]{\Why{#1}
  \crcr & 
  }
\newcommand{\StartRef}[1]{\hspace*{-1.5em}(\ref{#1}) \refsto
  \@ifnextchar[{\@StartRefCommment}{}}
\def\@StartRefCommment[#1]{\mbox{#1}
  \crcr \noalign{\penalty\interdisplaylinepenalty}}
\makeatother

\newcommand{\IFF}{\ChainRel{\Leftrightarrow}}

\newcommand{\Equiv}{\ChainRel{\equiv}}

\newcommand{\Refsto}{\ChainRel{\refsto}}

\newcommand{\Equals}{\ChainRel{=}}

\makeatletter
\def\@setmcodes#1#2#3{{\count0=#1 \count1=#3
  \loop \global\mathcode\count0=\count1 \ifnum \count0<#2
  \advance\count0 by1 \advance\count1 by1 \repeat}}

\DeclareSymbolFont{italic}{OT1}{\rmdefault}{m}{it}
\let\mathit\undefined
\DeclareSymbolFontAlphabet{\mathit}{italic}
\edef\@tempa{\hexnumber@\symitalic}
\@setmcodes{`A}{`Z}{"7\@tempa41}
\@setmcodes{`a}{`z}{"7\@tempa61}
\makeatother

\usepackage[colorinlistoftodos,textwidth=2cm]{todonotes}

\definecolor{Aqua}{rgb}{0,1,1}

\newcounter{hours}
\newcounter{minutes}
\newcommand{\printtime}{%
  \ifthenelse{\value{hours}<10}{0}{}\thehours:%
  \ifthenelse{\value{minutes}<10}{0}{}\theminutes}

\newbox{\MyDate}
\savebox{\MyDate}{\draftonly{ (\today\ \printtime)}}

\title{Distributive Laws for Parallel Composition \\ in Rely-Guarantee Concurrency}
\titlerunning{Distributive laws for parallel\usebox{\MyDate}}
\author{
Ian J. Hayes\inst{1}\orcidID{0000-0003-3649-392X}
\and
Larissa A. Meinicke\inst{1}\orcidID{0000-0002-5272-820X}
}
\titlerunning{Distributive Laws for Parallel Composition\usebox{\MyDate}} 
\institute{School of Electrical Engineering and Computer Science, \\ 
The University of Queensland, Brisbane, Queensland 4072, Australia 
}
\authorrunning{I. J. Hayes and L. A. Meinicke\usebox{\MyDate}}

\begin{document}

\maketitle

\begin{abstract}
The rely/guarantee approach supports the stepwise development of concurrent programs.
Our goal is to develop a theory for reasoning algebraically about concurrent programs in a rely/guarantee style,
where rely and guarantee conditions are encoding as commands within our theory.
As for mathematics, distributive laws are essential for algebraic manipulation of concurrent programs.
In this paper we investigate distributive laws for parallel composition 
and show how these can be applied to rely/guarantee concurrency.
The most general distributive laws are only refinements in a single direction,
however, by restricting the form of the command being distributed,
one can devise stronger equality laws, 
which are applicable to guarantee commands as well as to suitable combinations of rely and guarantee commands.
Our approach is to develop the distributive laws in a more abstract synchronous atomic algebra,
and then apply them to an instance of that algebra supporting rely and guarantee commands.
The theory has been formalised in Isabelle/HOL along with proofs of the lemmas presented here.
\end{abstract}

\section{Introduction}\labelsect{introduction}

\paragraph{Rely/guarantee concurrency.}

The rely-guarantee approach to reasoning about the correctness of shared memory concurrent programs
\cite{Jones81d,Jones83a,Jones83b} 
improved on its predecessors by providing a compositional approach,
that is, composite commands are reasoned about in terms of similar properties on their components,
as in Hoare logic \cite{Hoare69a}.
The main innovation was to introduce rely and guarantee conditions.
A rely condition, $r$, represents an assumption that 
the interference imposed on a thread by concurrently running threads is bounded by $r$, 
a binary relation on program states.
A guarantee condition, $g$, represents a commitment to ensure that
the interference imposed by a thread on threads running concurrently with it is bounded by $g$,
also a binary relation on program states.
Consider the following parallel version of Eratosthenes' sieve to determine the prime numbers up to $N$.
It consists of a set of parallel threads indexed over $i$ between 2 and $\lfloor \sqrt{N}\rfloor$,
each of which removes all multiples of its $i$ within the range $2 \upto N$,
where $mults\,i = \{ k * i \mid k \spot 2 \leq k \}$.
We use the predicative notation, $\Set{P}$,
for the set of states in which $P$ holds,
and $\Rel{R}$ for the set of pairs of states such that $R$ holds,
where a variable $v$ is interpreted as its value in the before state and $v'$ as its value in the after state.
\begin{align}
  \Spre{s = 2 \upto N} \Seq 
  \parallel_{i \in 2 \upto \lfloor\sqrt{N}\rfloor} 
   \left(\begin{array}{l}
    \Rrely{s' \subseteq s} \together {} \\
    \Rguar{s' \subseteq s \land s - s' \subseteq mults\,i} \together {} \\
    \Rpost{mults\,i \inter s' = \{\}}
 \end{array}\right)
\end{align}
The rely and guarantee conditions are encoded as commands $\rely{r}$ and $\guar{g}$ (see \refdef{guar} and \refdef{rely} below),
which can be combined to form a specification using the weak conjunction operation $\together$, 
explained in \refsect{CRA}.
The precondition $\Set{s = 2 \upto N}$ assumes $s$ is initially the full set of integers between 2 and $N$.
The rely condition $\Rel{s' \subseteq s}$ assumes all state-to-state transitions made by the environment of the thread
only remove elements from $s$.
The guarantee condition $\Rel{s' \subseteq s \land s - s' \subseteq mults\,i}$ 
ensures not only that the thread only removes elements from $s$ but it only removes multiples of $i$.
The postcondition $\Rel{mults\,i \inter s' = \{\}}$ ensures that, on termination, $s$ does not contain any multiples of $i$.
The combined postconditions of all the threads ensures that no multiples of any of the indices are left in $s$,
i.e.\ all composite numbers in the range have been removed,
and the guarantees of all the threads ensure that only multiples are removed,
i.e.\ only composites are removed.
Hence the final set is the prime numbers up to $N$.

\paragraph{Distributive laws.}

In mathematics, distributive laws such as, 
$x * (y + z) = (x * y) + (x * z)$, 
are essential for algebraic manipulation of formulae ---  the same is true for programs.
In earlier work \cite{2024MeinickeHayesDistributiveLaws} we developed distributive laws for relies and guarantees,
such as,
\begin{align}
  \guar{g} \together (c_1 \Seq c_2) & = (\guar{g} \together c_1) \Seq (\guar{g} \together c_2) \labelprop{guar-seq} \\
  \rely{r} \together (c_1 \Seq c_2) & = (\rely{r} \together c_1) \Seq (\rely{r} \together c_2) \labelprop{rely-seq} 
\end{align}
as well as distribution over fixed iteration, $c^i$ and finite iteration, $c^*$.
In the current paper we extend the distributive laws to distribute guarantees over parallel composition,
\begin{align}
  \guar{g} \together (c_1 \parallel c_2) & = (\guar{g} \together c_1) \parallel (\guar{g} \together c_2) \labelprop{guar-par} 
\end{align}
but note that we do not have an equivalent law for relies,
\begin{align}\color{purple}
  \rely{r} \together (c_1 \parallel c_2) = (\rely{r} \together c_1) \parallel (\rely{r} \together c_2) \labelprop{rely-par} 
\end{align}
because the rely on the left is an assumption about interference on the complete parallel composition 
generated by threads other than $c_1$ and $c_2$,
while the relies on the right represent interference generated by the other thread as well.
However, if $g \subseteq r$, which ensures the guarantees on each side of the parallel ensure the relies on the other side,
\begin{align}
  \rely{r} \together \guar{g} \together (c_1 \parallel c_2) = (\rely{r} \together \guar{g} \together c_1) \parallel (\rely{r} \together \guar{g} \together c_2) ,  \labelprop{rely-guar-par}
\end{align}
in particular, the evolution invariant command \cite{ColletteJones00a} defined as,
\begin{align}
  \evolve{r} \defs \rely{r} \together \guar{r},    \labeldef{evolve}
\end{align}
distributes over a parallel composition because $r \subseteq r$.

\paragraph{Defining the guarantee command.}

To define guarantees (and below relies) we make use of an approach devised by Peter Aczel \cite{Aczel83}
that distinguishes between transitions made by a thread --- its program transitions --- and
transitions made by all other threads running in concurrently with the thread --- its environment transitions.
The command, $\cpstep{g}$, allows a single program transition satisfying the binary relation between states $g$, 
and the command, $\cestep{r}$, allows a single environment transition satisfying $r$, 
and hence, $\cpstep{g} \nondet \cestep{r}$, which is a non-deterministic choice between the two, allows one transition or the other.
The guarantee command, $\guar{g}$, allows any program transitions satisfying $g$ and any environment transitions whatsoever,
and hence its basis command for a single transition is, $\cpstep{g} \nondet \cestep{\universalrel}$, 
where $\universalrel$ is the universal relation between states.
A guarantee command allows any sequence of zero or more such transitions, including infinite sequences.
The command $c^{\omega}$ represents the iteration of the command $c$ zero or more times, including infinitely many times,
and hence a guarantee command can be defined as 
the possibly infinite iteration of the basis atomic command $(\cpstep{g} \nondet \cestep{\universalrel})$.
\begin{align}
  \guar{g} \defs \Om{(\cpstep{g} \nondet \cestep{\universalrel})}    \labeldef{guar}
\end{align}

\paragraph{Defining the rely command.}

The command, $\Abort$, represents catastrophic irrecoverable failure,
i.e.\ Dijkstra's abort command \cite{Dijkstra75}.
It allows any behaviour whatsoever.
In the refinement calculus \cite{BackWright98,Morgan90}, 
an assert command, $\Pre{p}$, represents an assumption that the state is in the set of states $p$:
if the initial state is in $p$ it is a no-op, otherwise it aborts, as represented by $\Abort$.
Similarly, a rely command, $\rely{r}$, represents an assumption that all environment transitions satisfy $r$,
and hence it aborts if its environment makes a transition not satisfying $r$.
It is defined in terms of a possibly infinite iteration,
where its basis command allows any program transitions 
as well as environment transitions satisfying $r$
but it aborts if its environment makes a transition not satisfying $r$,
(i.e.\ in the complement $\overline{r}$ of the relation $r$).
\begin{align}
  \rely{r} \defs \Om{(\cpstep{\universalrel} \nondet \cestep{r} \nondet \cestep{\overline{r}} \Seq \Abort)}  \labeldef{rely}
\end{align}

\paragraph{Atomic commands.}

A command of the form $\cpstep{g} \nondet \cestep{r}$ is an \emph{atomic} command 
because it can make only a single transition.
In order to prove generic distribution laws, we make use of abstract atomic commands,
for which we use the names $\Ata$ and $\Atb$.
In earlier work \cite{2024MeinickeHayesDistributiveLaws}, we have shown that for an atomic command, $\Ata$,
\begin{align}
  \Om{\Ata} \together (c_1 \Seq c_2) = (\Om{\Ata} \together c_1) \Seq (\Om{\Ata} \together c_2) \labelprop{atom-iter-seq}
\end{align}
and hence \refprop{guar-seq} is an instance of this law.
As is shown in \Theorem*{par-distrib-omega} below, 
if $\Ata = \Ata \parallel \Ata$, 
\begin{align}
  \Om{\Ata} \together (c_1 \parallel c_2) = (\Om{\Ata} \together c_1) \parallel (\Om{\Ata} \together c_2) . \labelprop{iter-conj-par}
\end{align}
The proof of refinement from left to right is straightforward because from $\Ata = \Ata \parallel \Ata$
one can show $\Om{\Ata} \refsto \Om{\Ata} \parallel \Om{\Ata}$ and hence,
\begin{align*}
  \Om{\Ata} \together (c_1 \parallel c_2) \refsto (\Om{\Ata} \parallel \Om{\Ata}) \together (c_1 \parallel c_2) \refsto (\Om{\Ata} \together c_1) \parallel (\Om{\Ata} \together c_2) 
\end{align*}
where the final refinement step follows by the weak interchange law between $\together$ and $\parallel$,
\begin{align}
  (c_1 \parallel c_2) \together (d_1 \parallel d_2) \refsto (c_1 \together d_1) \parallel (c_2 \together d_2) . \labelax{weak-interchangex}
\end{align}
However, the reverse direction of the refinment for \refprop{iter-conj-par} requires a considerably more complex proof (see \refsect{dist-par}).
If $\Ata$ is of the form $\cpstep{g} \nondet \cestep{r}$,
then, as is shown in \reflem*{par-idempotent-aczel} below, 
$\Ata = \Ata \parallel \Ata$ if and only if $g \subseteq r$.
This holds for the guarantee command as $g \subseteq \universalrel$, for any $g$,
and hence \refprop{guar-par} is valid.

\paragraph{Pseudo-atomic commands.}

The rely command is not an iteration of a (pure) atomic command
but its basis is of the form $\Ata \nondet \Atb \Seq \Abort$, where $\Ata$ and $\Atb$ are atomic commands.
We refer to commands of this form as \emph{pseudo-atomic} commands.
For a pseudo atomic command, $\Patx$, if $\Patx = \Patx \parallel \Patx$, then 
\begin{align}
  \Om{\Patx} \together (c_1 \parallel c_2) = (\Om{\Patx} \together c_1) \parallel (\Om{\Patx} \together c_2) . \labelprop{iter-pseudo-conj-par}
\end{align}
In fact, at a suitable level of abstraction \refprop{iter-conj-par} and \refprop{iter-pseudo-conj-par} are the same lemma.
If $\Patx$ is of the form $\cpstep{g} \nondet \cestep{r} \nondet (\cpstep{h} \nondet \cestep{s}) \Seq \Abort$,
then as is shown in \reflem*{par-idempotent-aczel},
$\Patx = \Patx \parallel \Patx$, if and only if $g \subseteq r$ and $g \inter s \subseteq h \subseteq r \union s$.
Note that this condition is \underline{not} satisfied by a rely command 
(and hence \refprop{rely-par} is not valid) 
because $g$ is $\universalrel$, $r$ is $r$, $h$ is $\emptyset$ and $s$ is $\overline{r}$ 
but $\universalrel \not\subseteq r$, 
(unless $r$ is $\universalrel$, noting that $\rely{\universalrel}$ makes no assumption about its environment).
The composite command, $\guar{g} \together \rely{r}$, is equivalent to 
$\Om{(\cpstep{g} \nondet \cestep{r} \nondet \cestep{\overline{r}} \Seq \Abort)}$,
and this composite command distributes over parallel if $g \subseteq r$ and
$g \inter \overline{r} \subseteq \emptyset \subseteq r \union \overline{r}$,
which reduces to $g \subseteq r$ because that implies $g \inter \overline{r} = \emptyset$
and hence \refprop{rely-guar-par} is valid.

\refsect{CRA} gives the details of our concurrent refinement algebra in terms of weak quantales and biquantales
augmented with tests and atomic commands.
\refsect{dist-par} gives our main results for distribution laws over parallel.
\refsect{pseudo-atomic} extends to laws to handle pseudo-atomic commands.
\refsect{aczel-algebra} gives the Aczel algebra instance of the theory
in which rely and guarantee commands can be defined,
and calculates the conditions under which the distribution laws apply for relies and guarantees.
Related work is covered within paragraphs labeled ``Related work'' within the exposition and in the conclusions.

\section{Concurrent refinement algebra}\labelsect{CRA}

Our concurrent refinement algebra (CRA) is based on a lattice of commands with sequential, parallel and weak conjunction operations.
The theory is structured using quantale and biquantale structures as discussed in detail in \cite{2024RAMiCSrestructuring}.
A model of commands as sets of Aczel traces may be found in \cite{DaSMfaWSLwC}
but the presentation here focuses on their algebra.

\paragraph{Naming and syntactic precedence conventions.}
We use 
$c$ and $d$ for commands;
$p$ for sets of program states;
$g$, $q$ and $r$ for binary relations between program states;
$\Ata$ and $\Atb$ for atomic commands;
and
$\Patx$ for pseudo-atomic commands.
Subscripted versions of the above names follow the same convention.
Unary operations and function application have higher syntactic precedence than binary operations.
For binary operations, non-deterministic choice ($\nondet$) has the lowest precedence, 
and sequential composition ($\Seq$) has the highest precedence.
We use parentheses to resolve all other syntactic ambiguities.

\subsubsection*{Lattice of commands}

We make use of a wide spectrum language with both specification and programming language commands. 
The set of commands $\Command$ forms a complete distributive lattice where
\begin{itemize}
\item
the lattice partial order, $c \refsto d$, corresponds to refinement of commands, (i.e.\ $c$ is implemented by $d$),
\item
the lattice supremum, $\Nondet C$, corresponds to non-deterministic choice over a set of commands $C$ ---
it can perform any of the behaviours of commands within $C$,
\item
the lattice infimum, $\Meet C$, corresponds to strong conjunction ---
it can perform a behaviour only if all commands within $C$ can perform that behaviour,
\item
the bottom element of the lattice, $\Magic$, corresponds to the everywhere infeasible command,
where a command is infeasible in a state $\sigma$ if it can neither terminate nor abort nor make a transition in state $\sigma$,
and
\item
the top of the lattice, $\Abort$, corresponds to Dijkstra's abort command \cite{Dijkstra75} that has every possible behaviour.
\end{itemize}
\begin{lemmax*}[refine-choice]{\cite{BackWright98}}
If $\forall d \in D \spot \exists c \in C \spot c \refsto d$ then, $\Nondet C \refsto \Nondet D$.
\end{lemmax*}

\subsubsection*{Weak quantales}

Quantales form an algebraic structure that combines an associative operation $\odot$ with a complete lattice,
such that $\odot$ distributes from both the left and right over arbitrary suprema
\cite{2018StruthQuantales}.
In our context we need to weaken the distributive laws \refax{Nondet-distrib-op} and \refax{op-distrib-Nondet}
to only hold for non-empty $C$ because we include the irrecoverable aborting command $\Abort$ 
(see the related work discussion below).

\begin{definitionx}[weak quantale]
A \emph{weak quantale}, $(Q, \odot)$ consists of a complete lattice, $Q$, with an associative binary operation $\odot$ on $Q$
where for all $C \subseteq Q$ and $d \in Q$ the following two axioms hold.
\begin{align}
  (\Nondet_{c \in C} c) \odot d & = (\Nondet_{c \in C} c \odot d) & \mbox{if } C \neq \{\}  \labelax{Nondet-distrib-op} \\
  d \odot (\Nondet_{c \in C} c) & = (\Nondet_{c \in C} d \odot c) & \mbox{if } C \neq \{\}  \labelax{op-distrib-Nondet} 
\end{align}
If there exists a neutral element $\eta \in Q$ such that for all $c \in Q$, $c \odot \eta = \eta = \eta \odot c$, 
then $Q$ forms a \emph{unital weak quantale}, $(Q,\odot, \eta)$.
\end{definitionx}
For binary non-deterministic choice, $c \nondet d \defs \Nondet\{c,d\}$, the following two properties 
can be derived from \refax{Nondet-distrib-op} and \refax{op-distrib-Nondet}.
\begin{align}
  (c_1 \nondet c_2) \odot d & = c_1 \odot d \nondet c_2 \odot d \labelprop{nondet-distrib-op} \\
  c \odot (d_1 \nondet d_2) & = c \odot d_1 \nondet c \odot d_2 \labelprop{op-distrib-nondet}
\end{align}

\begin{definitionx}[abort-strict]
An operation, $\odot$, is abort strict if for all commands, $c$,
\begin{align}
  \Abort \odot c = \Abort .  \labelax{abort-strict}
\end{align}
\end{definitionx}

\begin{definitionx}[sequential quantale]
A \emph{sequential quantale}, $(Q, \Seq, \Nil)$, is a unital weak quantale for which sequential composition is \emph{abort strict} \refax{abort-seq}
and \refax{Nondet-distrib-op} holds for empty $C$ \refax{magic-seq}. 
\begin{align}
  \Abort \Seq c & = \Abort & \mbox{for all } c\in \Command \labelax{abort-seq} \\
  \Magic \Seq c & = \Magic & \mbox{for all } c \in \Command \labelax{magic-seq}
\end{align}
\end{definitionx}
Our theory includes three instances of unital weak quantales:
\begin{itemize}
\item
$(\Command, \Seq, \Nil)$ forms a sequential quantale 
where the neutral element (or unit) of sequential composition is $\Nil$,
\item
$(\Command,\parallel,\Skip)$ forms a unital weak quantale
where $\parallel$ is commutative and abort strict \refax{abort-strict}
and the neutral element of $\parallel$ is the command $\Skip$,
and
\item
$(\Command,\together, \Chaos)$ forms a unital weak quantale
where $\together$ is commutative, idempotent and abort strict \refax{abort-strict}
and the neutral element of $\together$ is the command $\Chaos$ that allows any non-aborting behaviour.
\end{itemize}
One difference between weak conjunction $\together$ and strong conjunction $\meet$
is that $\together$ is abort strict \refax{abort-strict} so that, 
$\Abort \together c = \Abort$,
whereas for strong conjunction, 
$\Abort \meet c = c$.
Noting that a sequential quantale is also a unital weak quantale,
all the lemmas proven for unital weak quantales hold in each of the above three instances,
thus avoiding triplication of work if one did factor out the common weak quantale structure.

\begin{RelatedWork}
Concurrent Kleene Algebra (CKA) \cite{DBLP:journals/jlp/HoareMSW11} uses (strong) quantales
and hence \refax{Nondet-distrib-op} and \refax{op-distrib-Nondet} hold for empty $C$ in CKA.
Taking $\odot$ as sequential composition,
\refax{op-distrib-Nondet} gives $c \Seq \Magic = \Magic$ for any command $c$ because $\Nondet \emptyset = \Magic$.
However, to reason about termination, one needs to handle infinite computations 
and, for example, $\Inf{c} \Seq d = \Inf{c}$ holds for any command $d$, including $\Magic$.
In addition, the abort command is irrecoverable, and so we also have $\Abort \Seq c = \Abort$ for any command $c$, including $\Magic$.
Both these requirements are needed to handle practical reasoning about rely/guarantee concurrency
and hence rule out the use of CKA as it stands.
\end{RelatedWork}

A sequential quantale gives one enough structure to define iteration operations \cite{Wright04}.
For $c$ a command and $i$ a natural number we define fixed iteration (\refdef*{fixed-iter-zero}--\refdef*{fixed-iter-unfold1}),
finite iteration as the least fixed point ($\mu$) of the monotone function $(\lambda z \spot \Nil \nondet c \Seq z)$ \refdef{finite-iter},
possibly infinite iteration as the greatest fixed point ($\nu$) of the same function \refdef{iter},
and
infinite iteration as the greatest fixed point of the monotone function $(\lambda z \spot c \Seq z)$ \refdef{inf-iter}.
From their definitions as fixed points, we get 
unfolding laws for iteration \refprop{iter-unfold} and infinite iteration \refprop{inf-iter-unfold},
and induction rules for iteration \refprop{iter-induct} and infinite iteration \refprop{inf-iter-induct}.
A possibly infinite iteration can be split into its finite and infinite components \refprop{isolation} 
and a finite iteration can be split into a choice over the natural numbers of fixed iterations \refprop{finite-iter-Nondet}.
\begin{eqncolumns}
  c^0 & = & \Nil   \labeldef{fixed-iter-zero} \\
  c^{i+1}  & = & c \Seq c^i  \labeldef{fixed-iter-unfold1} \\
  \Fin{c} & \defs & \mu z \spot \Nil \nondet c \Seq z \labeldef{finite-iter} \\
  \Om{c} & \defs & \nu z \spot \Nil \nondet c \Seq z \labeldef{iter} \\
  \Inf{c} & \defs & \nu z \spot c \Seq z \labeldef{inf-iter} \\
  \Om{c} & = & \Nil \nondet c \Seq \Om{c} \labelprop{iter-unfold} 
\secondcolumn
  \Inf{c} & = & c^i \Seq \Inf{c} \labelprop{inf-iter-unfold} \\
  \Om{c} \Seq d \refsto z && \mbox{if}~d \nondet c \Seq z \refsto z  \labelprop{iter-induct} \\
  \Inf{c}  \refsto  z && \mbox{if}~c \Seq z \refsto z \labelprop{inf-iter-induct} \\
  \Om{c} & = & \Fin{c} \nondet \Inf{c} \labelprop{isolation} \\
  \Fin{c} & = & \Nondet_{i \in \nat} c^i \labelprop{finite-iter-Nondet}
\end{eqncolumns}

\subsubsection*{Weak biquantales}

In order to define the interaction between two operations $\odot$ and $\sync$ we make use of weak biquantales,
which are weaker versions of the structure defined in Concurrent Kleene Algebra (CKA) \cite{DBLP:journals/jlp/HoareMSW11}.
An important axiom of a biquantale is the weak interchange axiom \refax{weak-interchange},
which for $\odot$ as sequential composition and $\sync$ as $\parallel$ gives,
\begin{align}
  (c_1 \Seq c_2) \parallel (d_1 \Seq d_2) & \refsto (c_1 \parallel d_1) \Seq (c_2 \parallel d_2) \labelax{seq-interchange-par}
\end{align}
which allows a parallel composition of two sequential compositions to be implemented by
synchronising their first components in parallel and then synchronising their second components.
The left side has other behaviours, such as $c_1$ synchronising with all of $d_1$ and part of $d_2$ 
and then $c_2$ synchronising with the remainder of $d_2$, or vice versa.
\begin{definitionx}[weak biquantale]
A \emph{weak biquantale}, $(B, \odot, \eta, \sync, \gamma)$ consists of a complete distributive lattice, $B$, where
$(B,\odot,\eta)$ and $(B,\sync,\gamma)$ form unital weak quantales and
$\sync$ is commutative and the following three axioms hold.
\begin{align}
  (c_1 \odot c_2) \sync (d_1 \odot d_2) & \refsto (c_1 \sync d_1) \odot (c_2 \sync d_2) \labelax{weak-interchange} \\
  \eta \sync \eta & \refsto \eta \labelax{neut-sync-neut} \\
  \gamma & \refsto \gamma \odot \gamma \labelax{gamma-op-gamma}
\end{align}
\end{definitionx}
From \refax{weak-interchange} with $c_1$ and $d_2$ both $\eta$ and $c_2$ and $d_1$ both $\gamma$ 
one can deduce \refprop{gamma-to-neut},
and from that and \refax{neut-sync-neut} and \refax{gamma-op-gamma} 
one can deduce that \refprop{neut-sync-neut} and \refprop{gamma-op-gamma} hold (see \cite{2024RAMiCSrestructuring}).
\\[-2ex]\begin{minipage}{0.34\textwidth}
\begin{align}
  \gamma & \refsto \eta \labelprop{gamma-to-neut}
\end{align}
\end{minipage}%
\begin{minipage}{0.33\textwidth}
\begin{align}
  \eta \sync \eta & = \eta \labelprop{neut-sync-neut} 
\end{align}
\end{minipage}%
\begin{minipage}{0.33\textwidth}
\begin{align}
  \gamma \odot \gamma & = \gamma \labelprop{gamma-op-gamma}
\end{align}
\end{minipage}

\begin{definitionx}[sequential biquantale]
A \emph{sequential biquantale}, $(B, \Seq, \Nil, \sync, \gamma)$,
is a weak biquantale for which $(B,\Seq,\Nil)$ forms a sequential quantale and
$\sync$ is abort strict. 
\end{definitionx}

We introduce the following instances of biquantales,
\begin{itemize}
\item
$(\Command, \Seq, \Nil, \parallel, \Skip)$ is a sequential biquantale,
\item
$(\Command, \Seq, \Nil, \together, \Chaos)$ is a sequential biquantale,
for which $\together$ is idempotent,
\item
$(\Command, \Seq, \Nil, \meet, \Abort)$ is a weak biquantale,
and
\item
$(\Command, \parallel, \Skip, \together, \Chaos)$ is a weak biquantale,
for which $\parallel$ is commutative and $\together$ is idempotent.
\end{itemize}
Again all the lemmas for a weak biquantale are available for each of the above biquantales,
noting that a sequential biquantale is a weak biquantale.

\begin{RelatedWork}
Unlike CKA \cite{DBLP:journals/jlp/HoareMSW11}, our neutral element, $\Skip$, of $\parallel$ 
is not the same as the neutral element $\Nil$ of sequential composition
but we do have $\Skip \refsto \Nil$.
That allows one to use a synchronous parallel operation similar to parallel in Milner's SCCS \cite{Milner83},
rather than an interleaving parallel composition.
\end{RelatedWork}

\subsubsection*{Tests, assertions and preconditions}

We follow Kozen's approach for Kleene Algebra with Tests (KAT) \cite{kozen97kleene}
and distinguish a subset of commands, $Test$, that are tests.
The set $Test$ is the image of the injective homomorphism, $\cgd{}$, from sets of states to commands,
so that the test command, $\cgd{p}$, terminates from states within the set $p$ but is infeasible otherwise.
The least test is $\cgd{\emptyset} = \Magic$ is the test that always fails and
the greatest test is $\cgd{\universalset} = \Nil$ is the test that always succeeds; 
note the bold font for the latter $\Nil$.
Because sets of states form a Boolean algebra, so do the subset of test commands,
where the lattice operations $\nondet$ and $\meet$ correspond to disjunction and conjunction.

Like the sequential refinement calculus \cite{BackWright98,Morgan90}, 
we define an assertion command, $\Pre{p}$ \refdef{assert}, that for a set of states $p$
is a no-op if the initial state is in $p$ but aborts otherwise 
(i.e.\ if the initial state is in the complement, $\overline{p}$, of $p$).
For a command $c$, the assertion, $\pre{c}$ \refdef{pre},
characterises the of states from which $c$ does not immediately abort \refdef{notimmedabort}.
\begin{align}
  \Pre{p} & \defs \cgd{p} \nondet \cgd{\overline{p}} \Seq \Abort \labeldef{assert} \\
  \pre{p} & \defs \Nondet \{ \Pre{p} \mid p \spot c \refsto \Pre{p} \Seq c \} \labeldef{pre} \\
  \notimmedabort{c} & \equiv (\pre{c} = \Nil) \labeldef{notimmedabort}
\end{align}
Many properties are shared by $\parallel$ and $\together$
and hence we give those properties once using the abstract commutative synchronisation operation $\sync$,
which is assumed to be abort strict \refax{abort-strict} because both $\parallel$ and $\together$ are abort strict.
Test, assert and precondition commands satisfy the following properties.
\begin{eqncolumns}[58]
  \Nil \sync c_1 \Seq c_2 & = & (\Nil \sync c_1) \Seq (\Nil \sync c_2) \labelax{test-sync-seq} \\
  \cgd{p_1} \sync \cgd{p_2} & = & \cgd{(p \inter p_2)} \labelax{test-sync-test} \\
  c_1 \sync \Pre{p} \Seq c_2 & = & \Pre{p} \Seq (c_1 \sync c_2) \labelprop{assert-command-sync-command} 
\secondcolumn
  \Nil \meet c & \in & Test \labelprop{nil-meet-is-test} \\
  \Nil \together c & = & \Nil \parallel c \labelprop{nil-conj-is-nil-par} \\
  \pre{c} \Seq c & = & c \labelprop{pre-seq} \\
  \Magic \sync c & = & \pre{c} \Seq \Magic \labelprop{magic-sync-c}
\end{eqncolumns}

\begin{lemmax}[sync-distrib-Nondet]
\(
  d \sync \Nondet C = \pre{d} \Seq \Nondet_{c \in C} (d \sync c) . 
\)
\end{lemmax}

\begin{proof}
If $C$ is empty,
$d \sync (\Nondet \emptyset)
= d \sync \Magic
= \pre{d} \Seq \Magic
= \pre{d} \Seq \Nondet_{c \in \emptyset} (d \sync c) 
$
as $\Nondet \emptyset = \Magic$ and \refprop{magic-sync-c}.
For $C$ non-empty, 
by \refax{op-distrib-Nondet}, \refprop{pre-seq}, \refprop{assert-command-sync-command}, and \refax{op-distrib-Nondet},
$d \sync \Nondet C 
= (\Nondet_{c \in C} d \sync c)
= (\Nondet_{c \in C} (\pre{d} \Seq d \sync c))
= (\Nondet_{c \in C} \pre{d} \Seq (d \sync c))
= \pre{d} \Seq (\Nondet_{c \in C} d \sync c)
$.
\end{proof}

\subsubsection*{Atomic commands} \labelsect{atomic}

We distinguish a subset of atomic commands, $\Atomic$, representing indivisible commands.
The greatest atomic command, $\cstepd \defs \Nondet \Atomic$, can perform any atomic transition
and the least atomic command is $\Magic$, which cannot perform any transition.
We assume atomic commands cannot immediately abort \refax{atom-non-abort}.
Synchronisation of atomic commands gives an atomic command \refax{atom-sync-atom-is-atom},
but note that sequential composition of atomic commands does not give an atomic command.
An atomic command, $\Ata$, synchronised with $\Nil$ is infeasible \refprop{atomic-sync-nil}.
The biquantale interchange axiom \refax{weak-interchange} for sequential composition and $\sync$ is strengthened to an equality
if the initial commands, $\Ata_1$ and $\Ata_2$, are atomic \refax{sync-interchange-seq-atomic}.
The sequential composition, $\Ata \Seq \Magic$, turns any terminating behaviour of the atomic command $\Ata$ into an incomplete behaviour.
If the result equals $\Ata$ then $\Ata$ must have had no terminating behaviours, i.e.\ $\Ata$ is everywhere infeasible \refax{atomic-seq-magic}.
Essentially, it constrains atomic commands so that if they have a behaviour that becomes infeasible after a transition,
then they also must have the behaviour that terminates after that transition, 
i.e.\ if an atomic command makes a transition, it must then terminate after the transition.
An atomic command prefixing a non-empty infimum of commands can be distributed into the infimum \refax{atom-seq-Inf}.
\begin{eqncolumns}[35]
  \pre{\Ata} & = & \Nil \labelax{atom-non-abort} \\
  \Ata_1 \sync \Ata_2 & \in & \Atomic \labelax{atom-sync-atom-is-atom} \\
  \Nil \sync \Ata & = & \Magic \labelprop{atomic-sync-nil} 
\secondcolumn
  \Ata_1 \Seq c_1 \sync \Ata_2 \Seq c_2 & = & (\Ata_1 \sync \Ata_2) \Seq (c_1 \sync c_2) \labelax{sync-interchange-seq-atomic} \\
  \Ata \Seq \Magic \refsto \Ata & \implies & \Ata = \Magic   \labelax{atomic-seq-magic} \\
  \Ata \Seq (\Meet C) & = & (\Meet_{c \in C} \Ata \Seq c)  ~~~~\mbox{if } C \not= \{\}  \labelax{atom-seq-Inf} 
\end{eqncolumns}
The atomic commands form a weak biquantale, $(\Atomic, \parallel, \cestepd, \together, \cstepd)$, 
where for atomic commands 
$\cestepd$ is the neutral element for $\parallel$ and 
$\cstepd$ is the neutral element for $\together$.
At this level we can define the commands $\Skip \defs \Om{\cestepd}$ and $\Chaos \defs \Om{\cstepd}$.
The interchange axiom \refax{weak-interchange} for this biquantale
shows that $(\Ata_1 \parallel \Ata_2) \together (\Atb_1 \parallel \Atb_2)$ is refined by both
$(\Ata_1 \together \Atb_1) \parallel (\Ata_2 \together \Atb_2)$ and  $(\Ata_1 \together \Atb_2) \parallel (\Ata_2 \together \Atb_1)$,
where the latter also uses commutativity of $\parallel$.
We strengthen this relationship to an equality with the choice of the two refinements,
that is, for atomic commands $\Ata_1$, $\Ata_2$, $\Atb_1$ and $\Atb_2$,
\begin{align}
  (\Ata_1 \parallel \Ata_2) \together (\Atb_1 \parallel \Atb_2) = 
  ((\Ata_1 \together \Atb_1) \parallel (\Ata_2 \together \Atb_2)) \nondet ((\Ata_1 \together \Atb_2) \parallel (\Ata_2 \together \Atb_1)) . \labelprop{conj-par-interchange-atomic}
\end{align}
In \refsect{aczel-algebra} we show this property holds in the Aczel algebra in which atomic commands have program and environment transitions.
Synchronisation of a test command with a command starting with an atomic command is infeasible.
\begin{lemmax}[test-sync-atomic]
For test $t$ and atomic command $\Ata$, $t \sync \Ata \Seq c = \Magic$.
\end{lemmax}

\begin{proof}
By \refprop{atomic-sync-nil}, $\Nil \sync \Ata = \Magic$ and hence 
applying \refax{test-sync-seq} 
$\Nil \sync \Ata \Seq c 
= (\Nil \sync \Ata) \Seq (\Nil \sync c) 
= \Magic \Seq (\Nil \sync c) 
= \Magic$
because $\Magic$ is a left annihilator for sequential composition by \refax{magic-seq}.
By \refax{test-sync-test} for any test $t$, $t = t \sync \Nil$, and we have, 
$t \sync \Ata \Seq c 
= t \sync \Nil \sync \Ata \Seq c 
= t \sync \Magic
= \Magic
$.
\end{proof}

\subsubsection*{Unrolled form of a command}

For an arbitrary command, $c$, its behaviour can be to 
abort if the precondition of $c$ does not hold initially (as represented by the assert command $\pre{c}$),
then if it did not abort it can either terminate immediately (as characterised by the test, $\Nil \meet c$), 
or do some atomic command $\Ata$ and then behave as some continuation command $c'$, 
for $(\Ata,c')$ in some set $\C$ of pairs of atomic commands and arbitrary commands.
That means any command $c$ can be expressed in an ``unrolled form'',%
\footnote{For readers familiar with transition systems,
an atomic command can be thought of as a set of possible transitions.
The  test, $\Nil \meet c$, represents a set of states in which $c$ may terminate,
(i.e.\ final states)
and for $(\Ata,c') \in \C$, the command $c$ can do any transition in $\Ata$ and then behave as $c'$.
One difference from transition systems is that 
an atomic command, $\Ata$, groups together transitions with the same continuation $c'$.
In addition, $c$ may abort from states not satisfying the precondition of $c$.
}
that is, for all commands $c$,
\begin{eqnarray}
  \exists C \spot (C \subseteq \Atomic \times \Command) \land (c = \pre{c} \Seq ((\Nil \meet c) \nondet \Nondet_{(\Ata,c') \in \C} (\Ata \Seq c'))) . \labelax{unrolled-form}
\end{eqnarray}

{\color{purple}See \ifarxiv \reflemy{unrolled-sync}{unrolled-proofs} \else \cite{TODO}\fi for proofs of the following two lemmas.}
\begin{lemmax}[unrolled-atomic]
If $\sync$ is abort strict and the unrolled form of $c$ is $c =\pre{c} \Seq ((\Nil \meet c) \nondet \Nondet_{(\Ata,c') \in \C} (\Ata \Seq c'))$
then,
\begin{align*}
  \Ata_1 \Seq c_1 \sync c = \pre{c} \Seq \Nondet_{(\Ata,c') \in \C} ((\Ata_1 \sync \Ata) \Seq (c_1 \sync c')) .
\end{align*}
\end{lemmax}

\begin{lemmax}[unrolled-sync]
If $\sync$ is abort strict and the unrolled forms are
$c_1 = \pre{c_1} \Seq ((\Nil \meet c_1) \nondet \Nondet_{(\Ata_1,c_1') \in \C_1} (\Ata_1 \Seq c_1'))$ and
$c_2 = \pre{c_2} \Seq ((\Nil \meet c_2) \nondet \Nondet_{(\Ata_2,c_2') \in \C_2} (\Ata_2 \Seq c_2'))$,
\[
  c_1 \sync c_2 = \pre{c_1} \Seq \pre{c_2} \Seq ((\Nil \meet c_1 \meet c_2) \nondet \Nondet_{(\Ata_1,c_1') \in \C_1} \Nondet_{(\Ata_2,c_2') \in \C_2} ((\Ata_1 \sync \Ata_2) \Seq (c_1' \sync c_2'))).
\]
\end{lemmax}

For arbitrary commands one can show, $\Fin{c_1} \sync \Fin{c_2} \refsto \Fin{(c_1 \sync c_2)}$
but for atomic commands this can be strengthened to an equality --- see \cite[Lemma 11]{FMJournalAtomicSteps} for a proof.
\begin{lemmax}[atomic-finite-iter-merge]
For atomic commands $\Ata_1$ and $\Ata_2$,
$\Fin{\Ata_1} \sync \Fin{\Ata_2} = \Fin{(\Ata_1 \sync \Ata_2)}$ .
\end{lemmax}

\section{Distribution over parallel}\labelsect{dist-par}

For a given atomic command $\Ata$, if 
\(
  \Fin{\Ata} \together (c_1 \parallel c_2) = (\Fin{\Ata} \together c_1) \parallel (\Fin{\Ata} \together c_2) ,
\)
holds for all $c_1$ and $c_2$ then, taking $c_1$ and $c_1$ to
both be the atomic unit, $\cstepd$, of $\together$, we must have that $\Ata$ is idempotent with
respect to parallel composition, i.e. $\Ata = \Ata \parallel \Ata$
because $\Fin{\Ata} \together \cstepd = \Ata$ and 
$\cstepd \parallel \cstepd = \cstepd$ by \refprop{gamma-op-gamma} for the atomic biquantale for $\parallel$ and $\together$.
The property $\Ata = \Ata \parallel \Ata$ is also a sufficient
condition for the distribution property to hold, and we develop a
proof of this below as \Theorem*{par-distrib-finite-iter}.
If $\Ata = \Ata \parallel \Ata$,
then $\Ata$ distributes over a parallel composition of atomic commands.
\begin{lemmax}[par-distrib-atomic]
For atomic commands $\Ata$, $\Ata_1$ and $\Ata_2$, if $\Ata = \Ata \parallel \Ata$, 
\[
  \Ata \together (\Ata_1 \parallel \Ata_2) = (\Ata \together \Ata_1) \parallel (\Ata \together \Ata_2) .
\]
\end{lemmax}

\begin{proof}
Using the assumption,
then \refprop{conj-par-interchange-atomic},
\(
  \Ata \together (\Ata_1 \parallel \Ata_2)
 = 
  (\Ata \parallel \Ata) \together (\Ata_1 \parallel \Ata_2)
 = 
  ((\Ata \together \Ata_1) \parallel (\Ata \together \Ata_2)) \nondet ((\Ata \together \Ata_2) \parallel (\Ata \together \Ata_1))
 = 
  (\Ata \together \Ata_1) \parallel (\Ata \together \Ata_2) 
\),
because $\parallel$ is commutative and $\nondet$ is idempotent.
\end{proof}

\begin{theoremx}[par-distrib-fixed-iter]
For an atomic command $\Ata$, 
if $\Ata = \Ata \parallel \Ata$, 
\[
  \Ata^i \together (c_1 \parallel c_2) = (\Ata^i \together c_1) \parallel (\Ata^ i \together c_2) .
\]
\end{theoremx}

\begin{proof}
By \refax{unrolled-form}
we can assume that there exist sets of pairs $\C_1$ and $\C_2$ of atomic commands and commands
such that 
$c_1 = \pre{c_1} \Seq ((\Nil \meet c_1) \nondet \Nondet_{(\Ata_1,c'_1) \in \C_1} (\Ata_1 \Seq  c'_1))$
and
$c_2 = \pre{c_2} \Seq ((\Nil \meet c_2) \nondet \Nondet_{(\Ata_2,c'_2) \in \C_2} (\Ata_2 \Seq  c'_2))$.
The proof is by induction on $i$. 
For $i = 0$, we have that $\Ata^0 = \Nil$ by \refdef{fixed-iter-zero}, 
and by \refprop{nil-conj-is-nil-par}, \refax{test-sync-test}, and \refprop{nil-conj-is-nil-par}.
\begin{align*}
\Nil \together (c_1 \parallel c_2)
= \Nil \parallel c_1 \parallel c_2
= (\Nil \parallel c_1) \parallel (\Nil \parallel c_2)
= (\Nil \together c_1) \parallel (\Nil \together c_2) .
\end{align*}
For the inductive case, we assume the property holds for $i$, and show it holds for $i+1$.
\begin{align*}&
  \Ata^{i+1} \together (c_1 \parallel c_2)
 \Equals*[by \refdef{fixed-iter-unfold1} and \reflem{unrolled-sync} with $\parallel$ for $\sync$ as $\parallel$ is abort strict]
  \Ata \Seq \Ata^i \together \pre{c_1} \Seq \pre{c_2} \Seq ((\Nil \meet c_1 \meet c_2) \nondet \Nondet_{(\Ata_1,c'_1) \in \C_1} \Nondet_{(\Ata_2,c'_2) \in \C_2} (\Ata_1 \parallel \Ata_2) \Seq (c'_1 \parallel c'_2))
 \Equals*[by \refprop{assert-command-sync-command} twice as $\together$ is abort strict and distribute $\Ata \Seq \Ata^i$ by \refprop{op-distrib-nondet} for $\together$]
  \pre{c_1} \Seq \pre{c_2} \Seq {} \\&
  ((\Ata \Seq \Ata^i \together (\Nil \meet c_1 \meet c_2)) \nondet \Ata \Seq \Ata^i \together (\Nondet_{(\Ata_1,c'_1) \in \C_1} \Nondet_{(\Ata_2,c'_2) \in \C_2} (\Ata_1 \parallel \Ata_2) \Seq (c'_1 \parallel c'_2)))
 \Equals*[as $\Ata \Seq \Ata^i \together (\Nil \meet c_1 \meet c_2) = \Magic$ by \reflem{test-sync-atomic} and \refprop{nil-meet-is-test}]
  \pre{c_1} \Seq \pre{c_2} \Seq (\Magic \nondet \Ata \Seq \Ata^i \together (\Nondet_{(\Ata_1,c'_1) \in \C_1} \Nondet_{(\Ata_2,c'_2) \in \C_2} (\Ata_1 \parallel \Ata_2) \Seq (c'_1 \parallel c'_2)))
 \Equals*[by \reflem{sync-distrib-Nondet} as $\pre{(\Ata \Seq \Ata^i)} = \pre{\Ata} = \Nil$ by \refax{atom-non-abort}]
  \pre{c_1} \Seq \pre{c_2} \Seq \Nondet_{(\Ata_1,c'_1) \in \C_1} \Nondet_{(\Ata_2,c'_2) \in \C_2} (\Ata \Seq \Ata^i \together (\Ata_1 \parallel \Ata_2) \Seq (c'_1 \parallel c'_2))
 \Equals*[by the reasoning below]
  \pre{c_1} \Seq \pre{c_2} \Seq 
  \Nondet_{(\Ata_1,c'_1) \in \C_1} \Nondet_{(\Ata_2,c'_2) \in \C_2} (((\Ata \together \Ata_1) \Seq (\Ata^i \together c'_1)) \parallel ((\Ata \together \Ata_2) \Seq (\Ata^i \together c'_2)))
 \Equals*[by \reflem{sync-distrib-Nondet} and \refprop{assert-command-sync-command} as $\parallel$ is abort strict \refax{abort-strict}]
  (\pre{c_1} \Seq \Nondet_{(\Ata_1,c'_1) \in \C_1} ((\Ata \together \Ata_1) \Seq (\Ata^i \together c'_1))) \parallel 
  (\pre{c_2} \Seq \Nondet_{(\Ata_2,c'_2) \in \C_2} ((\Ata \together \Ata_2) \Seq (\Ata^i \together c'_2)))
 \Equals*[by \reflem{unrolled-atomic} as $\together$ is abort strict \refax{abort-strict}; $\Ata \Seq \Ata^i = \Ata^{i+1}$ by \refdef{fixed-iter-unfold1}]
  (\Ata^{i+1} \together c_1) \parallel (\Ata^{i+1} \together c_2)
\end{align*}
The fifth proof step above holds as follows.
\begin{align*}&
  \Ata \Seq \Ata^i \together (\Ata_1 \parallel \Ata_2) \Seq (c'_1 \parallel c'_2)
 \Equals*[by atomic interchange \refax{sync-interchange-seq-atomic} for $\together$ as $\Ata$ and $(\Ata_1 \parallel \Ata_2)$ are atomic by \refax{atom-sync-atom-is-atom}]
  (\Ata \together (\Ata_1 \parallel \Ata_2)) \Seq (\Ata^i \together (c'_1 \parallel c'_2))
 \Equals*[by \reflem{par-distrib-atomic} as $\Ata = \Ata \parallel \Ata$ and inductive hypothesis]
  ((\Ata \together \Ata_1) \parallel (\Ata \together \Ata_2)) \Seq ((\Ata^i \together c'_1) \parallel (\Ata^i \together c'_2))
 \Equals*[by atomic interchange \refax{sync-interchange-seq-atomic} for $\parallel$ as $(\Ata \together \Ata_1)$ and $(\Ata \together \Ata_2)$ are atomic by \refax{atom-sync-atom-is-atom}]
  ((\Ata \together \Ata_1) \Seq (\Ata^i \together c'_1)) \parallel ((\Ata \together \Ata_2) \Seq (\Ata^i \together c'_2))
 \qedhere
\end{align*}
\end{proof}
\noindent
The above lemma can be promoted to finite iteration because $\Fin{\Ata} = \Nondet_{i \in \nat} \Ata^i$ by \refprop{finite-iter-Nondet}.
\begin{theoremx}[par-distrib-finite-iter]
If $\Ata$ is an atomic command such that $\Ata = \Ata \parallel \Ata$ then,
$\Fin{\Ata} \together (c_1 \parallel c_2) = (\Fin{\Ata} \together c_1) \parallel (\Fin{\Ata} \together c_2)$.
\end{theoremx}

\begin{proof}
We show refinement in both directions.
The refinement from left to right holds by \reflem{atomic-finite-iter-merge} as $\Ata = \Ata \parallel \Ata$ then 
weak interchange \refax{weak-interchange} of parallel and weak conjunction.
\begin{align*}&
  \Fin{\Ata} \together (c_1 \parallel c_2)
 = 
  (\Fin{\Ata} \parallel \Fin{\Ata}) \together (c_1 \parallel c_2)
 \refsto 
  (\Fin{\Ata} \together c_1) \parallel (\Fin{\Ata} \together c_2)
\end{align*}
The refinement from right to left holds as follows.
\begin{align*}&
  (\Fin{\Ata} \together c_1) \parallel (\Fin{\Ata} \together c_2)
 \Equals*[as $\Fin{\Ata} = \Nondet_{i \in \nat} \Ata^i$ by \refprop{finite-iter-Nondet} and distribute by \refax{Nondet-distrib-op} for $\together$ twice and $\parallel$ once] 
  (\Nondet_{j \in \nat} (\Ata^j \together c_1) \parallel (\Nondet_{k \in \nat}(\Ata^k \together c_2)))
 \Refsto*[as $\Nondet_{k \in \nat}(\Ata^k \together c_2) \refsto \Ata^j \together c_2$ by \reflem{refine-choice}] 
  \Nondet_{j \in \nat} ((\Ata^j \together c_1) \parallel (\Ata^j \together c_2)) 
 \Equals*[by \Theorem{par-distrib-fixed-iter} as $\Ata = \Ata \parallel \Ata$] 
  \Nondet_{j \in \nat} (\Ata^j \together (c_1 \parallel c_2))
 \Equals*[distribute by \refax{Nondet-distrib-op} for $\together$ and $\Fin{\Ata} = \Nondet_{j \in \nat} \Ata^j$ by \refprop{finite-iter-Nondet}] 
  \Fin{\Ata} \together (c_1 \parallel c_2)
 \qedhere
\end{align*}
\end{proof}
To extend this result to distributing a possibly infinite iteration, $\Om{\Ata}$, 
we need the machinery of limits 
of finite approximations, 
and super compact commands. 

\subsubsection*{Upper finite approximations}\labelsect{approx}

An finite upper approximation, $c \Approx i$, of a command, $c$, up to some natural number $i$ 
consists of all finite computations of $c$ of length less than $i$
plus computations of $c$ of length $i$ followed by abort.
For its definition we use the atomic neutral element, $\iota$, of the abort-strict operation $\sync$,
where $\iota$ corresponds to $\cestepd$ for $\parallel$ and $\cstepd$ for $\together$.
The notation $\Nondet_{j \in T}^{P\,j} c_j$ stands for the non-deterministic choice of $c_j$ 
for $j$ in $T$ such that $P\,j$ holds.
\begin{definitionx}[approx]
For $i \in \nat$, 
$c \Approx i \defs ((\Nondet_{k \in \nat}^{k < i} \iota^k) \nondet \iota^i \Seq \Abort)\sync c$.
\end{definitionx}
We have that $\iota^0 \Seq \Abort = \Abort$, and so the upper
finite approximation of $c$ up to length $0$ is $\Abort$ \refprop{approx-zero}.
For $i,j \in \nat$,  the approximation of $i$ initial atomic commands succeeded by a
command $c$ to length $i+j$ can be simplified by extracting the $i$
initial atomic commands, and reducing the approximation on $c$ to $j$ steps \refprop{seq-approx}.
Furthermore, approximations that exceed the length of the command have no effect \refprop{approx-more}.
Approximation distributes over non-deterministic choice \refprop{approx-nondet}.
Approximations of synchronisations of commands are equivalent if the approximations of the component commands are equivalent \refprop{approx-window-inf}.
These properties follow from \Definition{approx}.
\begin{align}
  c \Approx 0 & = \Abort \labelprop{approx-zero} \\
  (\Ata^i \Seq c) \Approx (i+j) & = \Ata^i \Seq (c \Approx j) \labelprop{seq-approx} \\
  \Ata^i \Approx (i+j) & = \Ata^i  &\mbox{if}~j > 0 \labelprop{approx-more} \\
  (c_1 \nondet c_2) \Approx i & = (c_1 \Approx i) \nondet (c_2 \Approx i) \labelprop{approx-nondet} \\
  (c_1 \sync c_2) \Approx i & = (d_1 \sync d_2) \Approx i &
  \mbox{if}~c_1 \Approx i = d_1 \Approx i ~\mbox{and}~ c_2 \Approx i = d_2 \Approx i  \labelprop{approx-window-inf} 
\end{align}
The approximation of an infinite number of iterations of $\Ata$ up to
$i$ steps simplifies to $\Ata^i \Approx i$, i.e. $\Ata^i \Seq \Abort$. 
\begin{lemmax}[atom-infinite-approx]
For an atomic command $\Ata$,
$\Inf{\Ata} \Approx i = \Ata^i \Approx i$.
\end{lemmax}
\begin{proof}
By unfolding the infinite iteration $i$ times using
\refprop{inf-iter-unfold}, and then applying \refprop{seq-approx}, and
\refprop{approx-zero}, 
applying \refprop{approx-zero} and \refprop{seq-approx} in reverse, and using
\refdef{fixed-iter-zero}, we have:
\begin{displaymath}
  \Inf{\Ata} \Approx i
= 
  (\Ata^i \Seq \Inf{\Ata}) \Approx i
= 
  \Ata^i \Seq (\Inf{\Ata} \Approx 0)
= 
  \Ata^i \Seq \Abort
= 
  \Ata^i \Seq (\Ata^0 \Approx 0)
= 
  (\Ata^i \Seq \Ata^0) \Approx i
= 
  \Ata^i \Approx i .
 \qedhere
\end{displaymath}
\end{proof}
From which we can show that all finite approximations of $\Om{\Ata}$ and $\Fin{\Ata}$ are equal. 
\begin{lemmax}[atom-omega-approx]
For an atomic command $\Ata$,
$\Om{\Ata} \Approx i = \Fin{\Ata} \Approx i$.
\end{lemmax}

\begin{proof}
By isolation \refprop{isolation},
distributing by \refprop{approx-nondet},
applying \reflem*{atom-infinite-approx},
reversing the distribution by \refprop{approx-nondet}
and then simplifying using \refprop{finite-iter-Nondet}, we have
\begin{math}
  \Om{\Ata} \Approx i
=
  (\Fin{\Ata} \nondet \Inf{\Ata}) \Approx i
=
  (\Fin{\Ata} \Approx i) \nondet (\Inf{\Ata} \Approx i)
=
  (\Fin{\Ata} \Approx i) \nondet (\Ata^i \Approx i)
=
  (\Fin{\Ata} \nondet \Ata^i) \Approx i
=
  \Fin{\Ata} \Approx i
\end{math}.
\end{proof}
Using \Theorem{par-distrib-finite-iter} we can show that if $\Ata = \Ata
\parallel \Ata$, then $\Om{\Ata}$ satisfies our desired distributivity
property for all finite approximations. 
\begin{lemmax}[atom-omega-par-approx]
For atomic command $\Ata$ such that, $\Ata = \Ata \parallel \Ata$, 
\[
  (\Om{\Ata} \together (c_1 \parallel c_2)) \Approx i = ((\Om{\Ata} \together c_1) \parallel (\Om{\Ata} \together c_2)) \Approx i
\]
\end{lemmax}

\begin{proof}
Using \refprop{approx-window-inf} with \reflem{atom-omega-approx} we have that for any $c$
\begin{align}
  (\Om{\Ata} \sync c)\Approx i & = (\Fin{\Ata} \sync c)\Approx i \labelprop{atom-omega-sync-approx} 
\end{align}
and so by \refprop{atom-omega-sync-approx}, \Theorem{par-distrib-finite-iter} using assumption $\Ata = \Ata \parallel \Ata$, and \refprop{approx-window-inf} with \refprop{atom-omega-sync-approx} twice:
\begin{math}
  (\Om{\Ata} \together (c_1 \parallel c_2)) \Approx i
= 
  (\Fin{\Ata} \together (c_1 \parallel c_2)) \Approx i
= 
  ((\Fin{\Ata} \together c_1) \parallel (\Fin{\Ata} \together c_2)) \Approx i 
= 
  ((\Om{\Ata} \together c_1) \parallel  (\Om{\Ata} \together c_2))  \Approx i
\end{math}.  
\end{proof}

\subsubsection*{Limits and limit-closure}\labelsect{limits}

The limit of the finite approximations of a command $c$ is the infimum of all finite approximations of $c$,
that is $\Meet_{i \in \nat} c \Approx i$.
\begin{definitionx}[limit-closed]
A command $c$ is limit closed if, $c = \Meet_{i \in \nat} c \Approx i$.
\end{definitionx}
Because the set of commands is assumed to form a complete lattice,
\reflem{fixed-point-continuity-limit} from fixed point theory applies,
and can be used to show that $\Om{\Ata}$ is limit closed, for any
atomic command $\Ata$. 
For a function $f \in \Command \rightarrow \Command$,
we define fixed iteration of $f$ via
$f^0\,z \defs z$ and
for $i \in \nat$, $f^{i+1}\,z \defs f(f^i\,z)$. 

\begin{lemmax}[fixed-point-continuity-limit]
If, for all non-empty sets of commands $C$, 
the monotone function $f\in \Command \rightarrow \Command$ satisfies, 
$f (\Meet C) = (\Meet_{c\in C} f\,c)$, then $\nu f = \Meet_{i\in \nat} f^i\,\Abort$.
\end{lemmax}

\begin{proof}
From \cite{DaveyPriestley02} using the assumption that commands form a complete lattice.
\end{proof}

\begin{lemmax}[approx-as-iteration]
If $f \defs (\lambda z \spot \Nil \nondet \Ata \Seq z)$, then $f^i\,\Abort = \Om{\Ata} \Approx i$.
\end{lemmax}

\begin{proof}
The proof is by induction on $i$.
For $i = 0$, $f^0\,\Abort = \Abort = \Om{\Ata} \Approx 0$ by \refprop{approx-zero}.
We assume the property for $i$ and show it holds for $i+1$ as follows:
$f^{i+1}\,\Abort 
= f(f^i\,\Abort) 
= \Nil \nondet \Ata \Seq f^i\,\Abort 
= \Nil \nondet \Ata \Seq (\Om{\Ata} \Approx i) 
= \Nil \nondet (\Ata \Seq \Om{\Ata}) \Approx (i+1) 
= (\Nil \nondet \Ata \Seq \Om{\Ata}) \Approx (i+1) 
= \Om{\Ata} \Approx (i+1) 
$
by the definition of $f$, 
the inductive hypothesis,
\refprop{seq-approx},
\refprop{approx-more},
\refprop{approx-nondet}
and
\refprop{iter-unfold}.
\end{proof}

\begin{lemmax}[omega-atom-limit-closed]
For atomic command $\Ata$, $\Om{\Ata}$ is limit closed.
\end{lemmax}

\begin{proof}
By \Definition{limit-closed}, we are required to show $\Om{\Ata} = \Meet_{i \in \nat} \Om{\Ata} \Approx i$.
which holds by \reflem*{fixed-point-continuity-limit} with $f$ the monotone function $(\lambda z \spot \Nil \nondet \Ata \Seq z)$
because $\nu f = \Om{\Ata}$ by \refdef{iter} and $f^i\,\Abort = \Om{\Ata} \Approx i$ by \reflem*{approx-as-iteration}.
The proviso for \reflem*{fixed-point-continuity-limit} is that $f(\Meet C) = \Meet_{c \in C} f\,c$ 
for any non-empty set of commands $C$,
which holds by \refax{atom-seq-Inf} and the fact that commands form a complete distributive lattice
because
$f(\Meet C) 
= \Nil \nondet \Ata \Seq (\Meet C)
= \Nil \nondet (\Meet_{c \in C} \Ata \Seq c)
= (\Meet_{c \in C} \Nil \nondet \Ata \Seq c)
= (\Meet_{c\in C} f~c)
$.
\end{proof}
Not all commands are limit-closed. For example, if
limit-closure was taken as an axiom, then both finite and possibly
infinite iterations of any atomic step $\Ata$ would be forced to
coincide by \reflem*{atom-omega-approx}:
$\Fin{\Ata} = \Meet_{i \in \nat} \Fin{\Ata} \Approx i = \Meet_{i \in  \nat} \Om{\Ata} \Approx i = \Om{\Ata}$.
Because we do not assume that all commands are limit closed, 
we cannot immediately infer from \reflem{atom-omega-par-approx} that
\refprop{iter-conj-par} holds if $\Ata = \Ata \parallel \Ata$.
To prove that distributivity property we require further
assumptions about the structure of our carrier set $\Command$.

\subsubsection*{Super-compact commands}\labelsect{super-compact}

\newcommand{\SuperCompactCommand}{\mathcal{D}}

The super-compact commands,
$\SuperCompactCommand \subseteq \Command$,
are those that are not immediately aborting and $\Nondet$-irreducible, i.e.\ they are the deterministic commands.
\begin{definitionx}[super-compact]
A command $y$ is \emph{super compact} if $y$ is not immediately aborting and for all non-empty sets of commands $C$,
\begin{align*}
  (\Nondet C \refsto y) \implies (\exists c \in C \spot c \refsto y) .
\end{align*}
\end{definitionx}
We make the additional assumption that the carrier set, $\Command$, is \emph{super-algebraic}.
\begin{definitionx}[super-algebraic]
The lattice of commands, $\Command$, is \emph{super-algebraic} if 
every command, $c$, can be written as 
a choice over a non-empty set of super compact commands preceded by is precondition \refprop{super-algebraic}.
\begin{align}
  \exists Y \spot  (\emptyset \neq Y \subseteq \SuperCompactCommand) & \land (c = \pre{c} \Seq \Nondet Y)  & \mbox{if $c \in \Command$} \labelprop{super-algebraic} 
\end{align}
\end{definitionx}
In our context we also assume that 
super-compact commands are limit-closed \refprop{super-compact-is-limit-closed},
super-compact commands are closed under synchronisation \refprop{super-compact-sync-super-compact},
and that the weak conjunction of a limit-closed command $c$ and a
super-compact command $y$ is super-compact if $c$ is not immediately aborting \refprop{limit-closed-together-super-compact}.
These properties hold for our trace model.
\begin{align}
  y & = \textstyle\Meet_{i \in \nat} y \Approx i  & \mbox{if $y\in \SuperCompactCommand$}  \labelprop{super-compact-is-limit-closed} \\
  y_1 \sync y_2 & \in \SuperCompactCommand &\mbox{if $y_1, y_2\in \SuperCompactCommand$} 
    \labelprop{super-compact-sync-super-compact} \\
  c \together y & \in \SuperCompactCommand  &\mbox{if $c$ is limit-closed, $\notimmedabort{c}$, $y \in \SuperCompactCommand$}
    \labelprop{limit-closed-together-super-compact}
\end{align}

\begin{lemmax}[par-distrib-omega-super-compact]
If $\Ata$ is an atomic command such that, $\Ata = \Ata \parallel \Ata$, 
and $y_1$ and $y_2$ are super-compact commands then 
\[
  \Om{\Ata} \together (y_1 \parallel y_2) = (\Om{\Ata} \together y_1) \parallel (\Om{\Ata} \together y_2) .
\]
\end{lemmax}

\begin{proof}
From \reflem{atom-omega-par-approx} and assumption $\Ata = \Ata \parallel \Ata$, 
it is sufficient to show that both 
$\Om{\Ata} \together (y_1 \parallel y_2)$
and
$(\Om{\Ata} \together y_1) \parallel (\Om{\Ata} \together y_2)$
are super-compact, and therefore, from \refprop{super-compact-is-limit-closed}, limit-closed. 
This follows from the assumptions on $y_1$ and $y_2$, the fact that $\Om{\Ata}$ is limit-closed by \reflem{omega-atom-limit-closed}, and not immediately aborting, and \refprop{super-compact-sync-super-compact} and \refprop{limit-closed-together-super-compact}.
\end{proof}

Using the assumption that the carrier set is super-algebraic, 
we can then extend \reflem*{par-distrib-omega-super-compact} to arbitrary commands
and thus show \refprop{iter-conj-par}.

\begin{theoremx}[par-distrib-omega]
If $\Ata$ is an atomic command such that, $\Ata = \Ata \parallel \Ata$, 
\[
  \Om{\Ata} \together (c_1 \parallel c_2) = (\Om{\Ata} \together c_1) \parallel (\Om{\Ata} \together c_2) .
\]
\end{theoremx}

\begin{proof}
From assumption \refprop{super-algebraic} we have that we can decompose $c_1 = \pre{c_1} \Seq \Nondet Y_1$ and
$c_2 = \pre{c_2} \Seq \Nondet Y_2$ into their super-compact components where $Y_1$ and $Y_2$ are non-empty subsets of $\SuperCompactCommand$.
\begin{align*}&
  \Om{\Ata} \together (c_1 \parallel c_2)
\Equals*[decompose $c_1 = \pre{c_1} \Seq \Nondet Y_1$ and $c_2 = \pre{c_2} \Seq \Nondet Y_2$]
  \Om{\Ata} \together (\pre{c_1} \Seq (\Nondet_{y_1 \in Y_1} y_1) \parallel \pre{c_2} \Seq (\Nondet_{y_2 \in Y_2} y_2))
\Equals*[distributing by \refprop{assert-command-sync-command}, \refax{Nondet-distrib-op} and \refax{op-distrib-Nondet} for $\parallel$ and \refax{op-distrib-Nondet} for $\together$ as $Y_1 \neq \emptyset$ and $Y_2 \neq \emptyset$]
  \pre{c_1}\Seq \pre{c_2} \Seq \Nondet_{y_1 \in Y_1} \Nondet_{y_2 \in Y_2} (\Om{\Ata} \together (y_1 \parallel y_2))
\Equals*[from \reflem{par-distrib-omega-super-compact}]
  \pre{c_1}\Seq \pre{c_2} \Seq \Nondet_{y_1 \in Y_1} \Nondet_{y_2 \in Y_2} (\Om{\Ata} \together y_1) \parallel (\Om{\Ata} \together y_2)
\Equals*[re-distribute by \refax{op-distrib-Nondet} and \refax{Nondet-distrib-op} for $\parallel$ then  \refax{op-distrib-Nondet} for $\together$ and \refprop{assert-command-sync-command} for $\parallel$ and $\together$]
(\Om{\Ata} \together \pre{c_1}\Seq (\Nondet_{y_1 \in Y_1} y_1)) \parallel
(\Om{\Ata} \together \pre{c_2} \Seq (\Nondet_{y_2 \in Y_2} y_2))
\Equals*[recompose $c_1 = \pre{c_1} \Seq \Nondet Y_1$ and
                   $c_2 = \pre{c_2} \Seq \Nondet Y_2$]
(\Om{\Ata} \together c_1) \parallel (\Om{\Ata} \together c_2)
\qedhere
\end{align*}
\end{proof}

\section{Pseudo-atomic commands}\labelsect{pseudo-atomic}

A \emph{pseudo-atomic} command, is a command, $\Patx$, that can be written in the form, 
$\Ata \nondet \Atb \Seq \Abort$, for some atomic commands $\Ata$ and $\Atb$.
Pseudo-atomic commands are closed under the basic operators of the algebra, except sequential composition.
\begin{lemmax}[pseudo-atomic-closed]
If $\Patx_1 = \Ata_1 \nondet \Atb_1 \Seq \Abort$ and $\Patx_2 = \Ata_2 \nondet \Atb_2 \Seq \Abort$ 
where $\Ata_1$, $\Atb_1$, $\Ata_2$ and $\Atb_2$ are atomic commands,
and $\sync$ is abort strict,
\begin{align}
  \Patx_1 \sync \Patx_2 & = (\Ata_1 \sync \Ata_2) \nondet (\Ata_1 \sync \Atb_2 \nondet \Atb_1 \sync \Ata_2 \nondet \Atb_1 \sync \Atb_2) \Seq \Abort \labelprop{pseudo-sync-closed}
\end{align}
and therefore because atomic commands are closed under 
$\sync$ and $\nondet$,
so are pseudo-atomic commands.
\end{lemmax}

\begin{proof}
We expand the forms of $\Patx_1$ and $\Patx_2$, and apply  \refprop{nondet-distrib-op}, \refprop{op-distrib-nondet}, \refax{sync-interchange-seq-atomic} 
and \refax{abort-strict} as $\sync$ is abort strict.
\begin{align*}&
  \Patx_1 \sync \Patx_2
 \Equals
  (\Ata_1 \nondet \Atb_1 \Seq \Abort) \sync (\Ata_2 \nondet \Atb_2 \Seq \Abort)
 \Equals
  (\Ata_1 \sync \Ata_2) \nondet (\Ata_1 \Seq \Nil \sync \Atb_2 \Seq \Abort) \nondet (\Atb_1 \Seq \Abort \sync \Ata_2 \Seq \Nil) \nondet (\Atb_1 \Seq \Abort \sync \Atb_2 \Seq \Abort)
 \Equals
  (\Ata_1 \sync \Ata_2) \nondet (\Ata_1 \sync \Atb_2 \nondet \Atb_1 \sync \Ata_2 \nondet \Atb_1 \sync \Atb_2) \Seq \Abort
 \qedhere
\end{align*}
\end{proof}

\begin{lemmax}[pseudo-atomic-eq-b]
If $\Ata_1 \nondet \Atb_1 \Seq \Abort = \Ata_2 \nondet \Atb_2 \Seq \Abort$ then $\Atb_1 = \Atb_2$.
{\color{purple}See \ifarxiv \reflemy{pseudo-atomic-eq-b}{pseudo-proofs} \else \cite{TODO}\fi for a proof.} 
\end{lemmax}

\begin{lemmax}[par-pseudo-atomic-expand]
If $\Patx$ is a pseudo-atomic command of the form $\Ata \nondet \Atb \Seq \Abort$ then,
$\Patx \parallel \Patx = (\Ata \parallel \Ata) \nondet ((\Ata \nondet \Atb) \parallel \Atb) \Seq \Abort$.
\end{lemmax}

\begin{proof}
By \refprop{pseudo-sync-closed},
$\Patx \parallel \Patx 
= (\Ata \parallel \Ata) \nondet (\Ata \parallel \Atb \nondet \Atb \parallel \Atb) \Seq \Abort
= (\Ata \parallel \Ata) \nondet ((\Ata \nondet \Atb) \parallel \Atb) \Seq \Abort .
$
\end{proof}

The following lemma gives an equivalent formulation of $\Patx = \Patx \parallel \Patx$ 
for $\Patx = \Ata \nondet \Atb \Seq \Abort$.
It has two provisos:
(i) $\Ata \meet \Atb = \Magic$, and
(ii) $\Ata \refsto \Ata \parallel \Ata$.
Because $\Ata \nondet \Atb \Seq \Abort = (\Ata \meet \lnot \Atb) \nondet \Atb \Seq \Abort$ for any $\Ata$ and $\Atb$,
(i) can be satisfied by replacing $\Ata$ by $\Ata \meet \lnot \Atb$ within $\Patx$.%
\footnote{This can be thought of as a canonical form for a pseudo-atomic command $\Patx$.}
Assumption (ii) holds in the Aczel algebra, as shown in \refcorollary{atomic-to-parallel-aczel} below.

\begin{lemmax}[pseudo-atomic-par-idempotent]
For a pseudo-atomic command $\Patx$ of the form $\Ata \nondet \Atb \Seq \Abort$,
if $\Ata \meet \Atb = \Magic$ and $\Ata \refsto \Ata \parallel \Ata$ then, 
\begin{align}
  \Patx = \Patx \parallel \Patx & \mbox{ if and only if } 
    (\Ata = \Ata \parallel \Ata) \land (\Atb = (\Ata \nondet \Atb) \parallel \Atb) . \labelprop{pseudo-atomic-par-idempotent}
\end{align}
\end{lemmax}

\begin{proof}
The reverse direction holds by \reflem{par-pseudo-atomic-expand}.
For the forward direction, $\Patx = \Patx \parallel \Patx$ is equivalent to
\begin{align}
  \Ata \nondet \Atb \Seq \Abort  = (\Ata \parallel \Ata) \nondet ((\Ata \nondet \Atb) \parallel \Atb) \Seq \Abort . \labelprop{expanded-equiv}
\end{align}
By \reflem{pseudo-atomic-eq-b}, this implies,
\begin{align}
  \Atb = (\Ata \nondet \Atb) \parallel \Atb  \labelprop{equal-b}
\end{align}
which is the second conjunct in \refprop{pseudo-atomic-par-idempotent}.
From \refprop{expanded-equiv} and \refprop{equal-b} we get $\Ata \nondet \Atb \Seq \Abort = (\Ata \parallel \Ata) \nondet \Atb \Seq \Abort$.
That implies $(\Ata \parallel \Ata) \nondet \Atb \Seq \Abort \refsto \Ata$,
and conjoining $\cstepd$ to both sides gives, $((\Ata \parallel \Ata) \nondet \Atb \Seq \Abort) \meet \cstepd \refsto \Ata$,
which distributing $\cstepd$ and simplifying gives, $(\Ata \parallel \Ata) \nondet \Atb \refsto \Ata$,
which with assumption $\Ata \meet \Atb = \Magic$ implies $\Ata \parallel \Ata \refsto \Ata$.
Combining this with the assumption that $\Ata \refsto \Ata \parallel \Ata$ gives $\Ata = \Ata \parallel \Ata$,
giving us the first conjunct in \refprop{pseudo-atomic-par-idempotent}.
\end{proof}

Property \refprop{conj-par-interchange-atomic} can be extended to allow pseudo-atomic commands on one side of the weak conjunction.
\begin{lemmax}[conj-par-interchange-pseudo-atomic]
For pseudo-atomic commands $\Patx_1$ and $\Patx_2$, and atomic commands $\Ata_1$ and $\Ata_2$,
\begin{eqnarray*}
  (\Patx_1 \parallel \Patx_2) \together (\Ata_1 \parallel \Ata_2) = 
  ((\Patx_1 \together \Ata_1) \parallel (\Patx_2 \together \Ata_2)) \nondet ((\Patx_1 \together \Ata_2) \parallel (\Patx_2 \together \Ata_1)) .
\end{eqnarray*}
{\color{purple}See \ifarxiv \reflemy{conj-par-interchange-pseudo-atomic}{pseudo-proofs} \else \cite{TODO}\fi for a proof.} 
\end{lemmax}

\begin{lemmax}[interchange-pseudo-atomic]
For pseudo-atomic commands $\Patx_1$ and $\Patx_2$, if $\sync$ is abort-strict,
\begin{align}
  \Patx_1 \Seq c_1 \sync \Patx_2 \Seq c_2 = (\Patx_1 \sync \Patx_2) \Seq (c_1 \sync c_2)
\end{align}
\end{lemmax}

\begin{proof}
We assume $\Patx_1 = \Ata_1 \nondet \Atb_1 \Seq \Abort$ and $\Patx_2 = \Ata_2 \nondet \Atb_2 \Seq \Abort$.
\begin{align*}&
  \Patx_1 \Seq c_1 \sync \Patx_2 \Seq c_2
 \Equals*[from the forms of $\Patx_1$ and $\Patx_2$ and distribute by \refprop{nondet-distrib-op} and apply \refax{abort-seq}]
  (\Ata_1 \Seq c_1 \nondet \Atb_1 \Seq \Abort) \sync (\Ata_2 \Seq c_2 \nondet \Atb_2 \Seq \Abort)
 \Equals*[distributing by \refprop{nondet-distrib-op} and \refprop{op-distrib-nondet}]
  (\Ata_1 \Seq c_1 \sync \Ata_2 \Seq c_2) \nondet (\Ata_1 \Seq c_1 \sync \Atb_2 \Seq \Abort) \nondet (\Atb_1 \Seq \Abort \sync \Ata_2 \Seq c_2) \nondet (\Atb_1 \Seq \Abort \sync \Atb_2 \Seq \Abort)
 \Equals*[by atomic interchange \refax{sync-interchange-seq-atomic} and $\sync$ is abort strict \refax{abort-strict}]
  (\Ata_1 \sync \Ata_2) \Seq (c_1 \sync c_2) \nondet (\Ata_1 \sync \Atb_2) \Seq \Abort \nondet (\Atb_1 \sync \Ata_2) \Seq \Abort \nondet (\Atb_1 \sync \Atb_2)\Seq \Abort
 \Equals*[distributing by \refprop{nondet-distrib-op} and $\Abort$ annihilates from the left \refax{abort-seq}]
  (\Ata_1 \sync \Ata_2) \Seq (c_1 \sync c_2) \nondet (\Ata_1 \sync \Atb_2 \nondet \Atb_1 \sync \Ata_2 \nondet \Atb_1 \sync \Atb_2) \Seq \Abort \Seq (c_1 \sync c_2)
 \Equals*[distributing by \refprop{nondet-distrib-op}]
  (\Ata_1 \sync \Ata_2  \nondet (\Ata_1 \sync \Atb_2 \nondet \Atb_1 \sync \Ata_2 \nondet \Atb_1 \sync \Atb_2) \Seq \Abort) \Seq (c_1 \sync c_2)
 \Equals*[by \reflem{pseudo-atomic-closed} and the forms of $\Patx_1$ and $\Patx_2$]
  (\Patx_1 \sync \Patx_2) \Seq (c_1 \sync c_2)
 \qedhere
\end{align*}
\end{proof}

Because pseudo-atomic commands satisfy the axioms for an atomic concurrent refinement algebra,
in particular, \refax{sync-interchange-seq-atomic} as shown in \reflem*{interchange-pseudo-atomic},
they satisfy the laws derived above but with pseudo-atomic commands in place of atomic commands.
\begin{theoremx}[par-distrib-pseudo-atomic]
For a pseudo-atomic command $\Patx$, 
the following all hold 
if $\Patx = \Patx \parallel \Patx$.
\begin{align}
  \Patx \together (\Ata_1 \parallel \Ata_2) & = (\Patx \together \Ata_1) \parallel (\Patx \together \Ata_2) \labelprop{pseudo-par-distrib-atomic} \\
  \Patx^i \together (c_1 \parallel c_2) & = (\Patx^i \together c_1) \parallel (\Patx^ i \together c_2) \labelprop{pseudo-par-distrib-fixed-iter} \\
  \Fin{\Patx} \together (c_1 \parallel c_2) & = (\Fin{\Patx} \together c_1) \parallel (\Fin{\Patx} \together c_2) \labelprop{pseudo-par-distrib-finite-iter} \\
  \Om{\Patx} \together (c_1 \parallel c_2) & = (\Om{\Patx} \together c_1) \parallel (\Om{\Patx} \together c_2) \labelprop{pseudo-par-distrib-iter} 
\end{align}
\end{theoremx}

\section{Application to the Aczel algebra}\labelsect{aczel-algebra}

The Aczel CRA is an instantiation of the Atomic CRA 
for which (pure) atomic commands are of the form, $\cpstep{g} \nondet \cestep{r}$, for some binary relations on program states $g$ and $r$.
Parallel composition is synchronous in the sense that it combines two atomic commands in parallel to give an atomic command,
similar to Milner's SCCS \cite{Milner83}. 
A program command in parallel with an environment command gives 
a program command on the intersection of their relations \refax{par-pgm-env}.
Two environment commands in parallel give an environment command on the intersection of their relations \refax{par-env-env}.
Two program commands in parallel do not synchronise (as represented by $\Magic$) \refax{par-pgm-pgm},
and hence program commands of two threads must interleave (by combining with environment commands of the other thread).
For Aczel atomic commands, weak conjunction ($\together$) corresponds to 
strong conjunction ($\meet)$ (\refprop*{cpstep-conj-cpstep}--\refprop*{cpstep-conj-cestep}).
\begin{eqncolumns}
  \cpstep{g} \parallel \cestep{r} & = & \cpstep{(g \inter r)}  \labelax{par-pgm-env} \\
  \cestep{g} \parallel \cestep{r} & = & \cestep{(g \inter r)} \labelax{par-env-env} \\
  \cpstep{g} \parallel \cpstep{r} & = & \Magic \labelax{par-pgm-pgm} 
\secondcolumn
  \cpstep{g} \together \cpstep{r} & = & \cpstep{(g \inter r)} \labelprop{cpstep-conj-cpstep} \\
  \cestep{g} \together \cestep{r} & = & \cestep{(g \inter r)} \labelprop{cestep-conj-cestep} \\
  \cpstep{g} \together \cestep{r} & = & \Magic \labelprop{cpstep-conj-cestep} 
\end{eqncolumns}
Note that $\cestepd = \cestep{\universalrel}$ is the atomic neutral element for $\parallel$ and
$\cstepd = \cpstep{\universalrel} \nondet \cestep{\universalrel}$ is the atomic neutral element for $\together$.
Two Aczel atomic commands are equivalent if and only if 
their program and environment relations are the same \refprop{pgm-env-equal}.
\begin{align}
  \cpstep{g_1} \nondet \cestep{r_1} = \cpstep{g_2} \nondet \cestep{r_2} & \iff g_1 = g_2 \land r_1 = r_2 \labelprop{pgm-env-equal} 
\end{align}
The following lemma shows that assumption \refprop{conj-par-interchange-atomic} holds for Aczel atomic commands.
\begin{lemmax}[conj-par-interchange-aczel]
For Aczel atomic commands $\Ata_1$, $\Ata_2$, $\Atb_1$ and $\Atb_2$,~~
\(
  (\Ata_1 \parallel \Ata_2) \together (\Atb_1 \parallel \Atb_2) = 
  ((\Ata_1 \together \Atb_1) \parallel (\Ata_2 \together \Atb_2)) \nondet ((\Ata_1 \together \Atb_2) \parallel (\Ata_2 \together \Atb_1)). 
\)
\end{lemmax}
The proof is a straightforward (if complex) expansion of both sides to an equivalent form.
{\color{purple}See \ifarxiv \reflemy{conj-par-interchange-aczel}{aczel-proofs} \else \cite{TODO} \fi.}
We calculate the condition under which $\Ata = \Ata \parallel \Ata$ for an Aczel atomic command $\Ata$.%
\begin{lemmax}[par-idempotent-aczel]
If $\Ata = \cpstep{g} \nondet \cestep{r}$, then $\Ata = \Ata \parallel \Ata \iff g \subseteq r$.
\end{lemmax}

\begin{proof}
We have,
\(
    \Ata \parallel \Ata
 = (\cpstep{g} \nondet \cestep{r}) \parallel (\cpstep{g} \nondet \cestep{r})
 = \cpstep{(g \inter r)} \nondet \cestep{r} ,
\)
by \refax{par-pgm-env}, \refax{par-env-env} and \refax{par-pgm-pgm},
which by \refprop{pgm-env-equal} equals $\cpstep{g} \nondet \cestep{r}$ 
if and only if $g = g \inter r$, that is, $g \subseteq r$.
\end{proof}

\paragraph{Application to rely/guarantee.}

A command, $\guarall{r} \defs \Om{(\cpstep{r} \nondet \cestep{r})}$, is referred to as a combined guarantee.
Three atomic commands that satisfy $\Ata = \Ata \parallel \Ata$ by \reflem*{par-idempotent-aczel} are:
\begin{itemize}
\item
the basis of a program guarantee \refdef{guar}, $\cpstep{g} \nondet \cestep{\universalrel}$, because $g \subseteq \universalrel$,
\item
the basis of a combined guarantee, $\cpstep{r} \nondet \cestep{r}$, because $r \subseteq r$,
and
\item
$\cestep{r}$ because $\emptyset \subseteq r$.
\end{itemize}
Hence by \Theorem{par-distrib-omega}, 
program guarantees \refprop{guar-par}, combined guarantees and $\Om{(\cestep{r})}$ 
all distribute over parallel composition.
Note that if $g \subseteq r$,  $\cpstep{g} \nondet \cestep{r} = (\cpstep{g} \nondet \cestep{\universalrel}) \together (\cpstep{r} \nondet \cestep{r})$,
and hence in the Aczel algebra,
any atomic command $\Ata$ of the form $\cpstep{g} \nondet \cestep{r}$ satisfying $\Ata = \Ata \parallel \Ata$ 
can be written as a conjunction of the basis of a program guarantee (for some $g$) and the basis of a combined guarantee (for some $r$),
so that $\Om{\Ata} = \guar{g} \together \guarall{r}$.
\begin{lemmax}[nondet-to-par-aczel]
For Aczel atomic commands $\Ata_1$ and $\Ata_2$, $\Ata_1 \nondet \Ata_2 \refsto \Ata_1 \parallel \Ata_2$.
{\color{purple}See \ifarxiv \reflemy{nondet-to-par-aczel}{aczel-proofs} \else \cite{TODO} \fi for a proof.}
\end{lemmax}
Choosing both $\Ata_1$ and $\Ata_2$ to be $\Ata$ in the above lemma gives the following corollary because $\Ata \nondet \Ata = \Ata$.
\begin{corollaryx}[atomic-to-parallel-aczel]
In the Aczel algebra, $\Ata \refsto \Ata \parallel \Ata$.
\end{corollaryx}

\begin{lemmax}[par-idempotent-pseudo-aczel]
If $\Patx = \Ata \nondet \Atb \Seq \Abort$ and
$\Ata = \cpstep{g} \nondet \cestep{r}$ and $\Atb = \cpstep{h} \nondet \cestep{s}$, 
then,
$\Patx = \Patx \parallel \Patx$
if and only if 
$g \subseteq r$ and $g \inter s \subseteq h \subseteq r \union s$.
{\color{purple}See \ifarxiv \reflemy{par-idempotent-pseudo-aczel}{aczel-proofs} \else \cite{TODO} \fi for a proof.}
\end{lemmax}

\paragraph{Application to rely/guarantee.}

The condition in \reflem*{par-idempotent-pseudo-aczel} holds for $\guar{g} \together \rely{r}$ provided $g \subseteq r$
because $\guar{g} \together \rely{r} = \Om{(\cpstep{g} \nondet \cestep{r} \nondet \cestep{\overline{r}} \Seq \Abort)}$,
and hence by \Theorem*{par-distrib-pseudo-atomic}, \refprop{rely-guar-par} holds
for $g$ as $g$, $r$ as $r$, $h$ as $\emptyset$ and $s$ as $\overline{r}$,
and $g \inter \overline{r} \subseteq \emptyset$ by the assumption $g \subseteq r$.
A special case of this is for an evolution command \refdef{evolve}, $\evolve{r}$,
that is, the command $\guar{r} \together \rely{r}$,
which satisfies the assumption because $r \subseteq r$.
The condition does not hold for a rely command \refdef{rely} 
(i.e.\ \refprop{rely-par} does not hold), 
where $g$ is $\universalrel$, $r$ is $r$, $h$ is $\emptyset$ and $s$ is $\overline{r}$
(because $\universalrel \inter \overline{r} \not\subseteq \emptyset$, unless $r$ is $\universalrel$).

\section{Conclusion}\labelsect{conclusion}

Our goal is to support the development of concurrent programs using the rely-guarantee approach.
The focus of this paper is on developing laws to distribute guarantees and relies over parallel composition.
Our intended applications, such as handling data refinement,
require the laws to be equalities rather than refinements in a single direction.
The approach of Concurrent Kleene Algebra (CKA) \cite{DBLP:journals/jlp/HoareMSW11} to handling the rely/guarantee concurrency
supports only weaker (single direction) laws and 
is not faithful to the original conception of Jones \cite{Jones81d}
because CKA requires threads to maintain their guarantee
even after their rely condition has been invalidated.
Furthermore, CKA only supports partial correctness and hence cannot handle termination arguments
\cite{2026RecursionWhileLoops_arxiv}.

Our approach to handling rely and pre conditions is to model their failure by the command abort,
which is irrecoverable.
To combine rely and guarantee commands to form specifications,
we introduced the weak conjunction operation, $c \together d$,
which corresponds to strong conjunction, $c \meet d$, unless either $c$ or $d$ aborts,
in which case $c \together d$ aborts.
To handle both infinite and aborting behaviours,
we need to make use of \emph{weak} quantales and biquantales, whereas CKA uses (strong) quantales.

A weak biquantale provides only a weak interchange law (i.e.\ a refinement in one direction).
By introducing atomic commands, we can strengthen the interchange law to an equality, if the initial commands are atomic,
in a manner similar to Milner's SCCS \cite{Milner83}.
This allows us to handle \emph{equality} distribution laws for iterations of atomic commands.
To distribute fixed ($\Ata^i$) and finite ($\Fin{\Ata}$) iteration we make use of the fact that a command can be unrolled 
to reveal its immediately aborting, immediately terminating and initial atomic transitions \refax{unrolled-form}.
To distribute possibly infinite iteration ($\Om{\Ata}$) we make use of the fact that $\Om{\Ata}$ is limit closed,
(i.e.\ it is the limit of its finite approximations) 
to extend the distribution law for $\Fin{\Ata}$ to $\Om{\Ata}$,
making use of the fact that commands can be decomposed into their super-compact (deterministic) components.

Because pseudo-atomic commands satisfy most of the same axioms as atomic commands,
the lemmas derived for atomic commands also apply to pseudo-atomic commands.
In the Isabelle theories this is handled by introducing a locale for the the atomic concurrent refinement algebra
and then instantiating that locale for pseudo atomics by showing they satisfy the axioms of the locale,
in particular, \reflem{interchange-pseudo-atomic} shows the atomic interchange axiom \refax{sync-interchange-seq-atomic}
holds for pseudo-atomic commands.
Developing the laws in the abstract synchronous algebra is considerably simpler than 
developing separate laws explicitly for guarantees and relies in the Aczel algebra
and allows the laws developed here to be applied in a wider range of contexts.

\paragraph{Acknowledgements.}
Our research has been supported by the 
Australian Research Council  
under their Discovery Program Grant 
DP190102142
in collaboration with Cliff Jones,
and
by funding from the 
Department of Defence, administered through the Advanced Strategic Capabilities Accelerator
grant \emph{Verifying Concurrent Data Structures for Trustworthy Systems}.
Megan Roxburgh completed the initial version of the Isabelle/HOL theories for distribution over parallel as an Honours thesis project.

\bibliographystyle{plainurl}
\bibliography{ms}

\newpage\appendix

\section{Unrolled command synchronisations}\labelapp{unrolled-proofs}

For this section we assume $\sync$ is abort strict, i.e.\ $\Abort \sync c = \Abort$.
\begin{lemmax}[unrolled-test]
For test $t$ and command $c$,
$t \sync c = \pre{c} \Seq (t \meet c)$.
\end{lemmax}

\begin{proof}
\begin{align*}&
  t \sync c
 \Equals*[using unrolled form \refax{unrolled-form} for $c$ for some $C$]
  t \sync \pre{c} \Seq ((\Nil \meet c) \nondet \Nondet_{(\Ata,c') \in \C} (\Ata \Seq c')) 
 \Equals*[by \refprop{assert-command-sync-command} as $\sync$ is abort strict]
  \pre{c} \Seq (t \sync ((\Nil \meet c) \nondet \Nondet_{(\Ata,c') \in \C} (\Ata \Seq c'))) 
 \Equals*[by distributing $t$ by \refprop{op-distrib-nondet} for $\sync$ and \reflem*{sync-distrib-Nondet} then \refax{test-sync-test} and \reflem*{test-sync-atomic}]
  \pre{c} \Seq (t \meet c)
 \qedhere
\end{align*}
\end{proof}

\begin{lemmay}[unrolled-atomic]
If 
$c = \pre{c} \Seq ((\Nil \meet c) \nondet \Nondet_{(\Ata,c') \in \C} (\Ata \Seq c'))$
for some set $\C$ of pairs of atomic commands and commands, 
i.e.\ $\C \subseteq \Atomic \times \Command$,
\begin{align}
  \Ata_1 \Seq c_1 \sync c = \pre{c} \Seq \Nondet_{(\Ata,c') \in \C} ((\Ata_1 \sync \Ata) \Seq (c_1 \sync c')) .
\end{align}
\end{lemmay}

\begin{proof}
Note that $\pre{(\Ata_1 \Seq c_1)} = \pre{a_1} = \Nil$ by \refax{atom-non-abort} as $\Ata_1$ is atomic.
\begin{align*}&
  \Ata_1 \Seq c_1 \sync c
 \Equals*[using unrolled form for $c$]
  \Ata_1 \Seq c_1 \sync \pre{c} \Seq ((\Nil \meet c) \nondet \Nondet_{(\Ata,c') \in \C} (\Ata \Seq c'))
 \Equals*[by \refprop{assert-command-sync-command} as $\sync$ is abort strict and distributing by \refprop{op-distrib-nondet} for $\sync$]
  \pre{c} \Seq ((\Ata_1 \Seq c_1 \sync (\Nil \meet c)) \nondet (\Ata_1 \Seq c_1 \Nondet_{(\Ata,c') \in \C} (\Ata \Seq c')))
 \Equals*[by \reflem{test-sync-atomic}, \refprop{nil-meet-is-test} and \reflem{sync-distrib-Nondet}]
  \pre{c} \Seq (\Magic \nondet \Nondet_{(\Ata,c') \in \C} (\Ata_1 \Seq c_1 \sync \Ata \Seq c'))
 \Equals*[by atomic interchange \refax{sync-interchange-seq-atomic}]
  \pre{c} \Seq \Nondet_{(\Ata,c') \in \C} ((\Ata_1 \sync \Ata) \Seq (c_1 \sync c'))
 \qedhere
\end{align*}
\end{proof}

\begin{lemmay}[unrolled-sync]
If 
$c_1 = \pre{c_1} \Seq ((\Nil \meet c_1) \nondet \Nondet_{(\Ata_1,c_1') \in \C_1} (\Ata_1 \Seq c_1'))$ and
$c_2 = \pre{c_2} \Seq ((\Nil \meet c_2) \nondet \Nondet_{(\Ata_2,c_2') \in \C_2} (\Ata_2 \Seq c_2'))$,
for some sets $\C_1$ and $\C_2$ of pairs of atomic commands and commands, 
\[
  c_1 \sync c_2 = \pre{c_1} \Seq \pre{c_2} \Seq ((\Nil \meet c_1 \meet c_2) \nondet \Nondet_{(\Ata_1,c_1') \in \C_1} \Nondet_{(\Ata_2,c_2') \in \C_2} ((\Ata_1 \sync \Ata_2) \Seq (c_1' \sync c_2'))).
\]
\end{lemmay}

\begin{proof}
Note that as the instances of the atomic commands $\Ata_1$ and $\Ata_2$ in the unrolled forms are not immediately aborting
and hence $\pre{\Ata_1} = \pre{\Ata_2} = \Nil$, 
and hence we also have  $\pre{\Ata_1 \Seq c_1'} = \pre{\Ata_2 \Seq c_2'} = \Nil$.
By \refax{test-sync-test} synchronisation of a pair of tests is the same as their meet and hence 
$(\Nil \meet c_1) \sync (\Nil \meet c_2) = \Nil \meet c_1 \meet c_2$.
\begin{align*}&
  c_1 \sync c_2
 \Equals*[unrolled forms of $c_1$ and $c_2$]
  \left(\pre{c_1} \Seq ((\Nil \meet c_1) \nondet \Nondet_{(\Ata_1,c_1') \in \C_1} (\Ata_1 \Seq c_1'))\right) \sync \\&
  \left(\pre{c_2} \Seq ((\Nil \meet c_2) \nondet \Nondet_{(\Ata_2,c_2') \in \C_2} (\Ata_2 \Seq c_2'))\right)
 \Equals*[by \refprop{assert-command-sync-command} twice as $\sync$ is abort strict \refax{abort-strict}]
  \pre{c_1} \Seq \pre{c_2} \Seq (((\Nil \meet c_1) \nondet \Nondet_{(\Ata_1,c_1') \in \C_1} (\Ata_1 \Seq c_1')) \sync ((\Nil \meet c_2) \nondet \Nondet_{(\Ata_2,c_2') \in \C_2} (\Ata_2 \Seq c_2')))
 \Equals*[distributing by \refprop{nondet-distrib-op} and \refprop{op-distrib-nondet} for $\sync$ and \reflem{sync-distrib-Nondet}]
  \pre{c_1} \Seq \pre{c_2} \Seq 
   \begin{array}[t]{l}
  (((\Nil \meet c_1) \sync (\Nil \meet c_2)) \nondet {} \\
   ~(\Nondet_{(\Ata_2,c_2') \in \C_2} (\Nil \meet c_1) \sync (\Ata_2 \Seq c_2')) \nondet {} \\
   ~(\Nondet_{(\Ata_1,c_1') \in \C_1} (\Ata_1 \Seq c_1') \sync (\Nil \meet c_2)) \nondet {} \\
   ~((\Nondet_{(\Ata_1,c_1') \in \C_1}  (\Ata_1 \Seq c_1')) \sync (\Nondet_{(\Ata_2,c_2') \in \C_2} (\Ata_2 \Seq c_2'))))
  \end{array}
 \Equals*[by \reflem{test-sync-atomic} twice and preconditions are $\Magic$]
  \pre{c_1} \Seq \pre{c_2} \Seq {} \\&
   ((\Nil \meet c_1 \meet c_2) \nondet \Magic \nondet
   ((\Nondet_{(\Ata_1,c_1') \in \C_1}  (\Ata_1 \Seq c_1')) \sync (\Nondet_{(\Ata_2,c_2') \in \C_2} (\Ata_2 \Seq c_2'))))
 \Equals*[as $\Magic$ is the identity of $\nondet$ and \reflem{sync-distrib-Nondet} twice]
  \pre{c_1} \Seq \pre{c_2} \Seq ((\Nil \meet c_1 \meet c_2) \nondet 
    \Nondet_{(\Ata_1,c_1') \in \C_1}  \Nondet_{(\Ata_2,c_2') \in \C_2} ((\Ata_1 \Seq c_1') \sync (\Ata_2 \Seq c_2')))
 \Equals*[by atomic interchange \refax{sync-interchange-seq-atomic}]
  \pre{c_1} \Seq \pre{t_2} \Seq ((\Nil \meet c_1 \meet c_2) \nondet 
    \Nondet_{(\Ata_1,c_1') \in \C_1}  \Nondet_{(\Ata_2,c_2') \in \C_2} ((\Ata_1 \sync \Ata_2) \Seq (c_1' \sync c_2')))
 \qedhere
\end{align*}
\end{proof}

\section{Proofs of pseudo-atomic lemmas}\labelapp{pseudo-proofs}

\begin{lemmax}[seq-abort-magic]
If $\Ata \refsto  \Ata \Seq \Abort$ then, $\Ata = \Magic$.
\end{lemmax}

\begin{proof}
Using the assumption, and the facts that abort is a left annihilator \refax{abort-seq} and $\Abort$ is the greatest element,
$\Ata \Seq \Magic \refsto \Ata \Seq \Abort \Seq \Magic = \Ata \Seq \Abort \refsto \Ata$,
and therefore by \refax{atomic-seq-magic}, $\Ata = \Magic$.
\end{proof}

\begin{lemmax}[atomic-to-atomic-abort]
If $\Ata \refsto \Atb \Seq \Abort$ then, $\Atb = \Magic$.
\end{lemmax}

\begin{proof}
From the assumption,
$\Ata \meet \Atb \Seq \Abort = \Atb \Seq \Abort$,
which implies $\Ata \meet \Atb = \Atb \Seq \Abort$,
and hence as $\Atb \refsto \Ata \meet \Atb$,
we have $\Atb \refsto \Atb \Seq \Abort$
and hence by \reflem{seq-abort-magic},
$\Atb = \Magic$.
\end{proof}

\begin{lemmax}[pseudo-atomic-to-atomic-abort]
If $\Ata_1 \nondet \Atb_1 \Seq \Abort \refsto \Atb_2 \Seq \Abort$ then, $\Atb_1 \refsto \Atb_2$.
\end{lemmax}

\begin{proof}
Weak conjoining $\lnot \Atb_1$ to both sides of the assumption gives,
$\lnot \Atb_1 \together (\Ata_1 \nondet \Atb_1 \Seq \Abort) \refsto \lnot \Atb_1 \together (\Atb_2 \Seq \Abort)$,
which simplifying gives,
$\lnot \Atb_1 \meet \Ata_1 \refsto (\lnot \Atb_1 \meet \Atb_2) \Seq \Abort$,
which by \reflem{atomic-to-atomic-abort} gives $\lnot \Atb_1 \meet \Atb_2 = \Magic$,
and so $\Atb_1 \refsto \Atb_2$.
\end{proof}

\begin{lemmay}[pseudo-atomic-eq-b]
If $\Ata_1 \nondet \Atb_1 \Seq \Abort = \Ata_2 \nondet \Atb_2 \Seq \Abort$ then $\Atb_1 = \Atb_2$.
\end{lemmay}

\begin{proof}
The assumption implies $\Ata_1 \nondet \Atb_1 \Seq \Abort \refsto \Atb_2 \Seq \Abort$,
which by \reflem{pseudo-atomic-to-atomic-abort} implies $\Atb_1 \refsto \Atb_2$.
By symmetry, we also get $\Atb_2 \refsto \Atb_1$.
Hence $\Atb_1 = \Atb_2$.
\end{proof}

\begin{lemmay}[conj-par-interchange-pseudo-atomic]
For pseudo-atomic commands $\Patx_1$ and $\Patx_2$, and atomic commands $\Ata_1$ and $\Ata_2$,
\begin{eqnarray*}
  (\Patx_1 \parallel \Patx_2) \together (\Ata_1 \parallel \Ata_2) = 
  ((\Patx_1 \together \Ata_1) \parallel (\Patx_2 \together \Ata_2)) \nondet ((\Patx_1 \together \Ata_2) \parallel (\Patx_2 \together \Ata_1)) .
\end{eqnarray*}
\end{lemmay}

\begin{proof}
For the proof we assume the pseudo-atomic commands are of the form 
$\Patx_1 = \Ata'_1 \nondet \Atb_1 \Seq \Abort$ and $\Patx_2 = \Ata'_2 \nondet \Atb_2 \Seq \Abort$.
For the application of \refax{sync-interchange-seq-atomic}, $\Ata_1 \parallel \Ata_2$ is atomic for atomic commands $\Ata_1$ and $\Ata_2$ by \refax{atom-sync-atom-is-atom}.
\begin{align*}&
  (\Patx_1 \parallel \Patx_2) \together (\Ata_1 \parallel \Ata_2)
 \Equals*[forms of $\Patx_1$ and $\Patx_2$, and \refprop{pseudo-sync-closed} as $\parallel$ is abort strict \refax{abort-strict}]
  ((\Ata'_1 \parallel \Ata'_2) \nondet ((\Ata'_1 \parallel \Atb_2) \nondet (\Atb_1 \parallel \Ata'_2) \nondet (\Atb_1 \parallel \Atb_2)) \Seq \Abort) \together (\Ata_1 \parallel \Ata_2)
 \Equals*[distributing by \refprop{nondet-distrib-op} for $\together$ and \refax{sync-interchange-seq-atomic} for $\together$ and $\together$ is abort strict \refax{abort-strict}]
  ((\Ata'_1 \parallel \Ata'_2) \together (\Ata_1 \parallel \Ata_2)) \nondet {} \\&
  (((\Ata'_1 \parallel \Atb_2) \together (\Ata_1 \parallel \Ata_2)) \nondet 
   ((\Atb_1 \parallel \Ata'_2) \together (\Ata_1 \parallel \Ata_2)) \nondet 
   ((\Atb_1 \parallel \Atb_2) \together (\Ata_1 \parallel \Ata_2))) \Seq \Abort
 \Equals*[by \refprop{conj-par-interchange-atomic} four times]
  ((\Ata'_1 \together \Ata_1) \parallel (\Ata'_2 \together \Ata_2)) \nondet ((\Ata'_1 \together \Ata_2) \parallel (\Ata'_2 \together \Ata_1)) \nondet {} \\&
  (((\Ata'_1 \together \Ata_1) \parallel (\Atb_2 \together \Ata_2)) \nondet ((\Ata'_1 \together \Ata_2) \parallel (\Atb_2 \together \Ata_1)) \nondet {} \\&
  ~((\Atb_1 \together \Ata_1) \parallel (\Ata'_2 \together \Ata_2)) \nondet ((\Atb_1 \together \Ata_2) \parallel (\Ata'_2 \together \Ata_1)) \nondet {} \\&
  ~((\Atb_1 \together \Ata_1) \parallel (\Atb_2 \together \Ata_2)) \nondet ((\Atb_1 \together \Ata_2) \parallel (\Atb_2 \together \Ata_1))) \Seq \Abort
 \Equals*[distributing by \refprop{nondet-distrib-op} and \refprop{op-distrib-nondet} for $\parallel$]
   ((\Ata'_1 \together \Ata_1) \nondet (\Atb_1 \Seq \Abort \together \Ata_1)) \parallel ((\Ata'_2 \together \Ata_2) \nondet (\Atb_2 \Seq \Abort \together \Ata_2)) \nondet {} \\&
   ((\Ata'_1 \together \Ata_2) \nondet (\Atb_1 \Seq \Abort \together \Ata_2)) \parallel ((\Ata'_2 \together \Ata_1) \nondet (\Atb_2 \Seq \Abort \together \Ata_1))
 \Equals*[distributing by \refprop{nondet-distrib-op} for $\together$]
   ((\Ata'_1 \nondet \Atb_1 \Seq \Abort) \together \Ata_1) \parallel ((\Ata'_2 \nondet \Atb_2 \Seq \Abort) \together \Ata_2) \nondet {} \\&
   ((\Ata'_1 \nondet \Atb_1 \Seq \Abort) \together \Ata_2) \parallel ((\Ata'_2 \nondet \Atb_2 \Seq \Abort) \together \Ata_1)
 \Equals*[forms of $\Patx_1$ and $\Patx_2$]
  ((\Patx_1 \together \Ata_1) \parallel (\Patx_2 \together \Ata_2)) \nondet ((\Patx_1 \together \Ata_2) \parallel (\Patx_2 \together \Ata_1))
 \qedhere
\end{align*}
\end{proof}

\section{Proofs of Aczel atomics lemmas}\labelapp{aczel-proofs}

\begin{lemmay}[nondet-to-par-aczel]
For Aczel atomic commands $\Ata_1$ and $\Ata_2$, $\Ata_1 \nondet \Ata_2 \refsto \Ata_1 \parallel \Ata_2$.
\end{lemmay}

\begin{proof}
There exist $g$, $r$, $h$ and $s$ such that, 
$\Ata_1 = \cpstep{g} \nondet \cestep{r}$ and $\Ata_2 = \cpstep{h} \nondet \cestep{s}$.
\begin{align*}&
  \Ata_1 \nondet \Ata_2 \refsto \Ata_1 \parallel \Ata_2
 \Equiv*[forms of $\Ata_1$ and $\Ata_2$]
  \cpstep{g} \nondet \cestep{r} \nondet \cpstep{h} \nondet \cestep{s} \refsto (\cpstep{g} \nondet \cestep{r}) \parallel (\cpstep{h} \nondet \cestep{s})
 \Equiv*[non-deterministic choice combines $\cpstep{}$ commands and combines $\cestep{}$ commands]
  \cpstep{(g \union h)} \nondet \cestep{(r \union s)} \refsto \cpstep{((g \inter s) \union (r \inter h))} \nondet \cestep{(r \inter s)}
 \Equiv*[refinement of $\cpstep{}/\cestep{}$ commands corresponds to containment of their relations]
  g \union h \supseteq (g \inter s) \union (r \inter h) \mbox{ and } r \union s \supseteq r \inter s 
\end{align*}
both of which hold.
\end{proof}

\begin{lemmay}[par-idempotent-pseudo-aczel]
If $\Patx = \Ata \nondet \Atb \Seq \Abort$ and
$\Ata = \cpstep{g} \nondet \cestep{r}$ and $\Atb = \cpstep{h} \nondet \cestep{s}$, 
then,
$\Patx = \Patx \parallel \Patx$
if and only if 
$g \subseteq r$ and $g \inter s \subseteq h \subseteq r \union s$.
\end{lemmay}

\begin{proof}
For Aczel atomic commands $\Ata \refsto \Ata \parallel \Ata$ by \reflem{nondet-to-par-aczel} with $\Ata_1$ and $\Ata_2$ both $\Ata$,
and without loss of generality one can assume $\Ata \meet \Atb = \Magic$,
and hence the two provisos for \reflem{pseudo-atomic-par-idempotent} hold.
By \reflem{par-pseudo-atomic-expand}
$\Patx \parallel \Patx 
= (\Ata \parallel \Ata) \nondet ((\Ata \nondet \Atb) \parallel \Atb) \Seq \Abort
$,
which by \reflem{pseudo-atomic-par-idempotent}
is equal to $\Patx$ iff both $\Ata = \Ata \parallel \Ata$ and $\Atb = (\Ata \nondet \Atb) \parallel \Atb$.
The former holds by \reflem{par-idempotent-aczel} iff $g \subseteq r$.
For the second we expand the right side using the definitions of $\Ata$ and $\Atb$.
\begin{align*}&
  (\Ata \nondet \Atb) \parallel \Atb
 \Equals
  (\cpstep{g} \nondet \cestep{r} \nondet \cpstep{h} \nondet \cestep{s}) \parallel (\cpstep{h} \nondet \cestep{s})
 \Equals
  (\cpstep{(g \union h)} \nondet \cestep{(r \union s)}) \parallel (\cpstep{h} \nondet \cestep{s})
 \Equals
  \cpstep{(((g \union h) \inter s) \union ((r \union s) \inter h))} \nondet \cestep{((r \union s) \inter s)}
 \Equals
  \cpstep{((g \inter s) \union (h \inter s) \union (r \inter h))} \nondet \cestep{s}
\end{align*}
By \refprop{pgm-env-equal} the latter is equal to $\Atb$, (i.e.\ $\cpstep{h} \nondet \cestep{s}$), 
if $(g \inter s) \union (h \inter s) \union (r \inter h) = h$,
which holds iff
\begin{align*}&
  (g \inter s) \union (h \inter (r \union s)) = h
 \IFF
  ((g \inter s) \union (h \inter (r \union s)) \subseteq h) \land (h \subseteq (g \inter s) \union (h \inter (r \union s)))
 \IFF
  (g \inter s \subseteq h) \land (h \inter (\overline{r \union s}) \subseteq g \inter s)
 \IFF
  (g \inter s \subseteq h) \land (h  \subseteq r \union s \union (g \inter s)
 \IFF
  (g \inter s \subseteq h) \land (h  \subseteq r \union s)
 \qedhere
\end{align*}
\end{proof}

\begin{lemmay}[conj-par-interchange-aczel]
For Aczel atomic commands $\Ata_1$, $\Ata_2$, $\Atb_1$ and $\Atb_2$,
\begin{eqnarray*}
  (\Ata_1 \parallel \Ata_2) \together (\Atb_1 \parallel \Atb_2) = 
  ((\Ata_1 \together \Atb_1) \parallel (\Ata_2 \together \Atb_2)) \nondet ((\Ata_1 \together \Atb_2) \parallel (\Ata_2 \together \Atb_1)). 
\end{eqnarray*}
\end{lemmay}

\begin{proof}
Because each of $\Ata_1$, $\Ata_2$, $\Atb_1$ and $\Atb_2$ is an atomic command,
for some $g_i$, $r_i$,  $h_i$ and $s_i$ they can be written as
$\Ata_1 = \cpstep{g_1} \nondet \cestep{r_1}$,
$\Ata_2 = \cpstep{g_2} \nondet \cestep{r_2}$,
$\Atb_1 = \cpstep{h_1} \nondet \cestep{s_1}$, and
$\Atb_2 = \cpstep{h_2} \nondet \cestep{s_2}$.
\begin{align*}
  \Ata_1 \parallel \Ata_2 
& =
  (\cpstep{g_1} \nondet \cestep{r_1}) \parallel (\cpstep{g_2} \nondet \cestep{r_2})
  =
  \cpstep{((g_1 \inter r_2) \union (r_1 \inter g_2))} \nondet \cestep{(r_1 \inter r_2)}
\\
  \Atb_1 \parallel \Atb_2 
 & =
  (\cpstep{h_1} \nondet \cestep{s_1}) \parallel (\cpstep{h_2} \nondet \cestep{s_2})
  =
  \cpstep{((h_1 \inter s_2) \union (s_1 \inter h_2))} \nondet \cestep{(s_1 \inter s_2)}
\end{align*}
Hence the left side of the lemma can be unrolled as follows.
\begin{align*}&
  (\Ata_1 \parallel \Ata_2 ) \together (\Atb_1 \parallel \Atb_2)
 \Equals
  (\cpstep{((g_1 \inter r_2) \union (r_1 \inter g_2))} \nondet \cestep{(r_1 \inter r_2)}) \together {} \\&
  (\cpstep{((h_1 \inter s_2) \union (s_1 \inter h_2))} \nondet \cestep{(s_1 \inter s_2)})
 \Equals
  \cpstep{(((g_1 \inter r_2) \union (r_1 \inter g_2)) \inter ((h_1 \inter s_2) \union (s_1 \inter h_2)))} \nondet \cestep{(r_1 \inter r_2 \inter s_1 \inter s_2)}
 \Equals
  \cpstep{((g_1 \inter r_2 \inter h_1 \inter s_2) \union (g_1 \inter r_2 \inter s_1 \inter h_2) \union {} \\&
               ~~~~~(r_1 \inter g_2 \inter h_1 \inter s_2) \union (r_1 \inter g_2 \inter s_1 \inter h_2))} \nondet \cestep{(r_1 \inter r_2 \inter s_1 \inter s_2)} \numberthis\labelprop{left-side-expansion}
\end{align*}
For the right side, we have,
\begin{align*}&
  (\Ata_1 \together \Atb_1) \parallel (\Ata_2 \together \Atb_2)
 \Equals
  ((\cpstep{g_1} \nondet \cestep{r_1}) \together (\cpstep{h_1} \nondet \cestep{s_1})) \parallel ((\cpstep{g_2} \nondet \cestep{r_2}) \together (\cpstep{h_2} \nondet \cestep{s_2}))
 \Equals
  (\cpstep{(g_1 \inter h_1)} \nondet \cestep{(r_1 \inter s_1)}) \parallel (\cpstep{(g_2 \inter h_2)} \nondet \cestep{(r_2 \inter s_2)})
 \Equals
  \cpstep{((g_1 \inter h_1 \inter r_2 \inter s_2) \union (r_1 \inter s_1 \inter g_2 \inter h_2))} \nondet \cestep{(r_1 \inter s_1 \inter r_2 \inter s_2)}
\end{align*}
and similarly,
\begin{align*}&
  (\Ata_1 \together \Atb_2) \parallel (\Ata_2 \together \Atb_1)
 \Equals
  \cpstep{((g_1 \inter h_2 \inter r_2 \inter s_1) \union (r_1 \inter s_2 \inter g_2 \inter h_1))} \nondet \cestep{(r_1 \inter s_1 \inter r_2 \inter s_2)}
\end{align*}
therefore, expanding the right side we get the following,
\begin{align*}&
  ((\Ata_1 \together \Atb_1) \parallel (\Ata_2 \together \Atb_2)) \nondet ((\Ata_1 \together \Atb_2) \parallel (\Ata_2 \together \Atb_1))
 \Equals
  (\cpstep{((g_1 \inter h_1 \inter r_2 \inter s_2) \union (r_1 \inter s_1 \inter g_2 \inter h_2))} \nondet \cestep{(r_1 \inter s_1 \inter r_2 \inter s_2)}) \nondet {} \\&
  (\cpstep{((g_1 \inter h_2 \inter r_2 \inter s_1) \union (r_1 \inter s_2 \inter g_2 \inter h_1))} \nondet \cestep{(r_1 \inter s_1 \inter r_2 \inter s_2)})
 \Equals
  (\cpstep{((g_1 \inter h_1 \inter r_2 \inter s_2) \union (r_1 \inter s_1 \inter g_2 \inter h_2) \union {} \\& ~~~~~~(g_1 \inter h_2 \inter r_2 \inter s_1) \union (r_1 \inter s_2 \inter g_2 \inter h_1))} \nondet 
  \cestep{(r_1 \inter s_1 \inter r_2 \inter s_2)})
 \Equals*[see expansion for the left side \refprop{left-side-expansion} above]
  (\Ata_1 \parallel \Ata_2 ) \together (\Atb_1 \parallel \Atb_2)
 \qedhere
\end{align*}
\end{proof}

\end{document}

\section{Old lemmas}

\begin{lemmay}[assert-command-sync-command]
If $\sync$ is abort strict, that is, $c \sync \Abort = \Abort$ for all $c$, then
$c_1 \sync \Assert{t} \Seq c_2 = \Assert{t} \Seq (c_1 \sync c_2)$.
\end{lemmay}

\begin{proof}
\begin{align*}&
  c_1 \sync \Assert{t} \Seq c_2
 \Equals*[case analysis as $\Nil = t \nondet \lnot t$ and distributing by \refprop{nondet-distrib-op} for $\Seq$]
  t \Seq (c_1 \sync \Assert{t} \Seq c_2) \nondet \lnot t \Seq (c_1 \sync \Assert{t} \Seq c_2) 
 \Equals*[by \refax{test-distrib-sync} twice]
  (t \Seq c_1 \sync t \Seq \Assert{t} \Seq c_2) \nondet (\lnot t \Seq c_1 \sync \lnot t \Seq \Assert{t} \Seq c_2) 
 \Equals*[by \refprop{test-assert} and $\lnot t \Seq \Assert{t} = \lnot t \Seq (\Nil \nondet \lnot t\Seq \Abort) = \lnot t \Seq \Nil \nondet \lnot t \Seq \Abort = \lnot t \Seq (\Nil \nondet \Abort) = \lnot t \Seq \Abort$]
  (t \Seq c_1 \sync t \Seq c_2) \nondet (\lnot t \Seq c_1 \sync \lnot t \Seq \Abort) 
 \Equals*[by \refax{test-distrib-sync} twice]
  t \Seq (c_1 \sync c_2) \nondet \lnot t \Seq (c_1 \sync \Abort) 
 \Equals*[by the assumption that $\sync$ is abort strict \refax{abort-strict}]
  t \Seq (c_1 \sync c_2) \nondet \lnot t \Seq \Abort 
 \Equals*[as for any $c$ and $t$,  $\Assert{t} \Seq c = t \Seq c \nondet \lnot t \Seq \Abort$ by \refprop{assert-seq}]
  \Assert{t} \Seq (c_1 \sync c_2)
 \qedhere
\end{align*}
\end{proof}

\begin{lemmay}[sync-atomic-fixed-iter]{\cite[Lemma 8]{FMJournalAtomicSteps}}
For $i \in \nat$, and atomic commands $\Ata_1$ and $\Ata_2$,
$\Ata_1^i \Seq c_1 \sync \Ata_2^i \sync c_2 = (\Ata_1 \sync \Ata_2)^i \Seq (c_1 \sync c_2)$.
\end{lemmay}

\begin{proof}
The proof is by induction on $i$.
For $i = 0$, using \refdef{fixed-iter-zero},
$\Ata_1^0 \Seq c_1 \sync \Ata_2^0 \Seq c_2 = c_1 \sync c_2 = (\Ata_1 \sync \Ata_2)^0 \Seq (c_1 \sync c_2)$.
For the inductive case, 
assume $\Ata_1^i \Seq c_1 \sync \Ata_2^i \sync c_2 = (\Ata_1 \sync \Ata_2)^i \Seq (c_1 \sync c_2)$,
and show,
\begin{align*}&
  \Ata_1^{i+1} \Seq c_1 \sync \Ata_2^{i+1} \sync c_2 
 \Equals*[unfolding the fixed iteration \refdef{fixed-iter-unfold1}]
  \Ata_1 \Seq \Ata_1^{i} \Seq c_1 \sync \Ata_2 \Seq \Ata_2^{i} \sync c_2 
 \Equals*[by \refax{sync-interchange-seq-atomic}]
  (\Ata_1 \sync \Ata_2) \Seq (\Ata_1^{i} \Seq c_1 \sync \Ata_2^{i} \sync c_2)
 \Equals*[by inductive hypothesis]
  (\Ata_1 \sync \Ata_2) \Seq (\Ata_1 \sync \Ata_2)^i \Seq (c_1 \sync c_2)
 \Equals*[folding the fixed iteration \refdef{fixed-iter-unfold1}]
  (\Ata_1 \sync \Ata_2)^{i+1} \Seq (c_1 \sync c_2)
 \qedhere
\end{align*}
\end{proof}

\begin{lemmay}[sync-atomic-finite-iter]{\cite[Lemma 9]{FMJournalAtomicSteps}}
For not immediately aborting atomic commands $\Ata_1$ and $\Ata_2$,
\(
    \Fin{\Ata_1} \Seq c_1 \sync \Fin{\Ata_2} \Seq c_2
 = \Fin{(\Ata_1 \sync \Ata_2)} \Seq ((c_1 \sync  \Fin{\Ata_2} \Seq c_2) \nondet (\Fin{\Ata_1} \Seq c_1 \sync c_2)).
\)
\end{lemmay}

\begin{proof}
\begin{align*}&
  \Fin{\Ata_1} \Seq c_1 \sync \Fin{\Ata_2} \Seq c_2
 \Equals*[by finite iteration decomposition \refprop{finite-iter-Nondet} twice and distributing by \refax{op-distrib-Nondet} for $\Seq$ twice]
  \Nondet_{i \in \nat} \Ata_1^i \Seq c_1 \sync \Nondet_{j \in \nat} \Ata_2^j \Seq c_2
 \Equals*[by \refax{op-distrib-Nondet} for $\sync$]
  \Nondet_{i \in \nat} \Nondet_{j \in \nat} \Ata_1^i \Seq c_1 \sync \Ata_2^j \Seq c_2
 \Equals*[splitting choice using \reflem{refine-choice} and $\Ata_2^j = \Ata_2^i \Seq \Ata_2^{j-i}$ for $i \leq j$]
  (\Nondet_{i \in \nat} \Nondet_{j \in \nat}^{i \leq j} \Ata_1^i \Seq c_1 \sync \Ata_2^i \Seq \Ata_2^{j-i} \Seq c_2) \nondet 
  (\Nondet_{i \in \nat} \Nondet_{j \in \nat}^{i \geq j} \Ata_1^j \Seq \Ata_1^{j-i} \Seq c_1 \sync \Ata_2^j \Seq c_2)
 \Equals*[renaming bound variables with $k = j-i$ and $m = i-j$; $k$ and $m$ ranging over $\nat$]
  (\Nondet_{i \in \nat} \Nondet_{k \in \nat} \Ata_1^i \Seq c_1 \sync \Ata_2^i \Seq \Ata_2^k \Seq c_2) \nondet 
  (\Nondet_{i \in \nat} \Nondet_{m \in \nat} \Ata_1^j \Seq \Ata_1^m \Seq c_1 \sync \Ata_2^j \Seq c_2)
 \Equals*[distributing for $\sync$ \refax{op-distrib-Nondet} and sequential \refax{op-distrib-Nondet}]
  (\Nondet_{i \in \nat} \Ata_1^i \Seq c_1 \sync \Ata_2^i \Seq  \Nondet_{k \in \nat} \Ata_2^k \Seq c_2) \nondet 
  (\Nondet_{i \in \nat} \Ata_1^j \Seq \Nondet_{m \in \nat} \Ata_1^m \Seq c_1 \sync \Ata_2^j \Seq c_2)
 \Equals*[by \reflem{sync-atomic-fixed-iter} and \refprop{finite-iter-Nondet}]
  (\Nondet_{i \in \nat} (\Ata_1 \sync \Ata_2)^i \Seq (c_1 \sync \Fin{\Ata_2} \Seq c_2)) \nondet 
  (\Nondet_{i \in \nat} (\Ata_1 \sync \Ata_2)^i \Seq (\Fin{\Ata_1} \Seq c_1 \sync c_2))
 \Equals*[distributing by \refprop{op-distrib-nondet} for $\Seq$ and \refprop{finite-iter-Nondet}]
  \Fin{(\Ata_1 \sync \Ata_2)} \Seq ((c_1 \sync  \Fin{\Ata_2} \Seq c_2) \nondet (\Fin{\Ata_1} \Seq c_1 \sync c_2))
 \qedhere
\end{align*}
\end{proof}
The above proof is based on that in \cite{FMJournalAtomicSteps} and
repeated here to indicate the general structure of such proofs.
We do not give the proofs for \reflem{sync-atomic-finite-infinite} and \reflem{sync-atomic-iter} below in detail (see \cite{FMJournalAtomicSteps}).

Needed?
\begin{lemmax}[finite-aborting-distrib]
If $\Ata = \Ata \parallel \Ata$ and $\Atb \parallel (\Ata \nondet \Atb) = \Atb$,
\[
  (\Fin{\Ata} \Seq \Atb \Seq \Abort \together c_1) \parallel (\Fin{\Ata} \Seq \Atb \Seq \Abort \together c_2) \refsto \Fin{\Ata} \Seq \Atb \Seq \Abort \together (c_1 \parallel c_2).
\]
\end{lemmax}

\begin{proof}
\begin{align*}&
  (\Fin{\Ata} \Seq \Atb \Seq \Abort \together c_1) \parallel (\Fin{\Ata} \Seq \Atb \Seq \Abort \together c_2)
 \Equals*[as $\Fin{\Ata} = \Nondet_{i \in \nat} \Ata^i$ by \refprop{finite-iter-Nondet} and distributing \refax{Nondet-distrib-op} for $\Seq$ and $\together$]
  (\Nondet_{j \in \nat} \Ata^j \Seq \Atb \Seq \Abort \together c_1) \parallel (\Nondet_{k \in \nat} \Ata^k \Seq \Atb \Seq \Abort \together c_2)
 \Equals*[distributing \refax{op-distrib-Nondet} for $\parallel$]
  \Nondet_{j \in \nat} \Nondet_{k \in \nat} ((\Ata^j \Seq \Atb \Seq \Abort \together c_1) \parallel (\Ata^k \Seq \Atb \Seq \Abort \together c_2))
 \Refsto*[restricting $k$ to just $j$ and $j+1$ and duplicating choices for $j$ except for $j=0$]
  \Nondet_{j \in \nat} (((\Ata^j \nondet \Ata^{j+1}) \Seq \Atb \Seq \Abort \together c_1) \parallel ((\Ata^j \nondet \Ata^{j+1}) \Seq \Atb \Seq \Abort \together c_2))
 \Refsto*[see below]
  \Nondet_{j \in \nat} (\Ata^j \Seq \Atb \Seq \Abort \together (c_1 \parallel c_2))
 \Equals*[distributing by \refax{Nondet-distrib-op} for both $\together$ and $\Seq$ and $\Nondet_{j \in \nat} \Ata^j = \Fin{\Ata}$ by \refprop{finite-iter-Nondet}]
  \Fin{\Ata} \Seq \Atb \Seq \Abort \together (c_1 \parallel c_2)
\end{align*}
For the second last step we show 
\[
 \forall j \in \nat \spot 
  ((\Ata^j \nondet \Ata^{j+1}) \Seq \Atb \Seq \Abort \together c_1) \parallel ((\Ata^j \nondet \Ata^{j+1}) \Seq \Atb \Seq \Abort \together c_2)
  \refsto \Ata^j \Seq \Atb \Seq \Abort \together (c_1 \parallel c_2)
\]
by induction in $j$. For $j=0$, we have $\Ata^j = \Nil$ and $\Ata^{j+1} = \Ata$, 
\begin{align*}&
  ((\Nil \nondet \Ata) \Seq \Atb \Seq \Abort \together c_1) \parallel ((\Nil \nondet \Ata) \Seq \Atb \Seq \Abort \together c_2)
 \Equals*[distributing by \refprop{nondet-distrib-op} for $\Seq$]
  ((\Atb \Seq \Abort \together c_1) \nondet (\Ata \Seq \Atb \Seq \Abort \together c_1)) \parallel ((\Atb \Seq \Abort \together c_2) \nondet (\Ata \Seq \Atb \Seq \Abort \together c_2))
 \Refsto*[distributing by \refprop{op-distrib-nondet} for $\parallel$ and removing one alternative]
  ((\Atb \Seq \Abort \together c_1) \parallel (\Atb \Seq \Abort \together c_2)) \nondet
  ((\Atb \Seq \Abort \together c_1) \parallel (\Ata \Seq \Atb \Seq \Abort \together c_2)) \nondet
  ((\Ata \Seq \Atb \Seq \Abort \together c_1) \parallel (\Atb \Seq \Abort \together c_2))
 \Equals*[by \reflem{unrolled-atomic} six times as $\together$ is abort strict \refax{abort-strict}]
  (\Assert{t_1'} \Seq \Nondet_{(\Ata_1,c'_1) \in \C_1} \Atb \Seq \Abort \together \Ata_1 \Seq c'_1) \parallel 
  (\Assert{t_2'} \Seq \Nondet_{(\Ata_2,c'_2) \in \C_2} \Atb \Seq \Abort \together \Ata_2 \Seq c'_2) \nondet {} \\&
  (\Assert{t_1'} \Seq \Nondet_{(\Ata_1,c'_1) \in \C_1} \Atb \Seq \Abort \together \Ata_1 \Seq c'_1) \parallel 
  (\Assert{t_2'} \Seq \Nondet_{(\Ata_2,c'_2) \in \C_2} \Ata \Seq \Atb \Seq \Abort \together \Ata_2 \Seq c'_2) \nondet {} \\&
  (\Assert{t_1'} \Seq \Nondet_{(\Ata_1,c'_1) \in \C_1} \Ata \Seq \Atb \Seq \Abort \together \Ata_1 \Seq c'_1) \parallel 
  (\Assert{t_2'} \Seq \Nondet_{(\Ata_2,c'_2) \in \C_2} \Atb \Seq \Abort \together \Ata_2 \Seq c'_2)
 \Equals*[by \refprop{assert-command-sync-command} distribute]
  \Assert{t_1'} \Seq \Assert{t_2'} \Seq\Nondet_{(\Ata_1,c'_1) \in \C_1} \Nondet_{(\Ata_2,c'_2) \in \C_2} \\&
   (\Atb \Seq \Abort \together \Ata_1 \Seq c'_1) \parallel (\Atb \Seq \Abort \together \Ata_2 \Seq c'_2) \nondet {} \\&
   (\Atb \Seq \Abort \together \Ata_1 \Seq c'_1) \parallel (\Ata \Seq \Atb \Seq \Abort \together \Ata_2 \Seq c'_2) \nondet {} \\&
   (\Ata \Seq \Atb \Seq \Abort \together \Ata_1 \Seq c'_1) \parallel (\Atb \Seq \Abort \together \Ata_2 \Seq c'_2)
 \Equals*[distribute and \refax{sync-interchange-seq-atomic} for $\together$, which is abort strict \refax{abort-strict}]
  \Assert{t_1'} \Seq \Assert{t_2'} \Seq \Nondet_{(\Ata_1,c'_1) \in \C_1} \Nondet_{(\Ata_2,c'_2) \in \C_2} \\&
   ((\Atb \together \Ata_1) \Seq \Abort \parallel (\Atb \together \Ata_2) \Seq \Abort) \nondet {} \\&
   ((\Atb \together \Ata_1) \Seq \Abort \parallel (\Ata \together \Ata_2) \Seq (\Atb \Seq \Abort \together c'_2)) \nondet {} \\&
   ((\Ata \together \Ata_1) \Seq (\Atb \Seq \Abort \together c'_1) \parallel (\Atb \together \Ata_2) \Seq \Abort)
 \Equals*[by \refax{sync-interchange-seq-atomic} for $\parallel$, which is abort strict \refax{abort-strict}]
  \Assert{t_1'} \Seq \Assert{t_2'} \Seq \Nondet_{(\Ata_1,c'_1) \in \C_1} \Nondet_{(\Ata_2,c'_2) \in \C_2} \\&
   (((\Atb \together \Ata_1) \parallel (\Atb \together \Ata_2)) \nondet 
    ((\Atb \together \Ata_1) \parallel (\Ata \together \Ata_2)) \nondet 
    ((\Ata \together \Ata_1) \parallel (\Atb \together \Ata_2))) \Seq \Abort
 \Equals*[distributing by \refprop{op-distrib-nondet} for $\parallel$]
  \Assert{t_1'} \Seq \Assert{t_2'} \Seq \Nondet_{(\Ata_1,c'_1) \in \C_1} \Nondet_{(\Ata_2,c'_2) \in \C_2} \\&
   (((\Atb \together \Ata_1) \parallel (\Atb \together \Ata_2 \nondet \Ata \together \Ata_2)) \nondet 
    ((\Atb \together \Ata_1 \nondet \Ata \together \Ata_1) \parallel (\Atb \together \Ata_2))) \Seq \Abort 
 \Equals*[distributing by \refprop{op-distrib-nondet} for $\together$]
  \Assert{t_1'} \Seq \Assert{t_2'} \Seq \Nondet_{(\Ata_1,c'_1) \in \C_1} \Nondet_{(\Ata_2,c'_2) \in \C_2} \\&
   (((\Atb \together \Ata_1) \parallel ((\Atb \nondet \Ata) \together \Ata_2)) \nondet 
    (((\Atb \nondet \Ata) \together \Ata_1) \parallel (\Atb \together \Ata_2))) \Seq \Abort
 \Equals*[by \reflem{conj-par-interchange-pseudo-atomic} as $\Atb \parallel (\Ata \nondet \Atb) = \Atb$]
  \Assert{t_1'} \Seq \Assert{t_2'} \Seq \Nondet_{(\Ata_1,c'_1) \in \C_1} \Nondet_{(\Ata_2,c'_2) \in \C_2} \Atb \Seq \Abort \together (\Ata_1 \parallel \Ata_2) \Seq (c_1' \parallel c_2')
 \Equals*[distribute and \refprop{assert-command-sync-command} twice]
  \Atb \Seq \Abort \together (\Assert{t_1'} \Seq (t_1 \nondet \Nondet_{(\Ata_1,c'_1) \in \C_1} (\Ata_1 \Seq c_1')) \parallel \Assert{t_2'} \Seq (t_2 \nondet \Nondet_{(\Ata_2,c'_2) \in \C_2} (\Ata_2 \Seq c_2')))
 \Equals*[contracting unrolled forms of $c_1$ and $c_2$]
  \Atb \Seq \Abort \together (c_1 \parallel c_2)
\end{align*}
For the inductive case, we assume the property for $j$ and show it for $j+1$.
\begin{align*}&
  ((\Ata^{j+1} \nondet \Ata^{j+2}) \Seq \Atb \Seq \Abort \together c_1) \parallel 
  ((\Ata^{j+1} \nondet \Ata^{j+2}) \Seq \Atb \Seq \Abort \together c_2)
 \Equals*[as $\Ata^{j+1} \nondet \Ata^{j+2} = \Ata \Seq (\Ata^j \nondet \Ata^{j+1})$]
  (\Ata \Seq (\Ata^{j} \nondet \Ata^{j+1}) \Seq \Atb \Seq \Abort \together c_1) \parallel 
  (\Ata \Seq (\Ata^{j} \nondet \Ata^{j+1}) \Seq \Atb \Seq \Abort \together c_2)
 \Equals*[unrolled forms of $c_1$ and $c_2$, \reflem{unrolled-atomic}, \refprop{assert-command-sync-command}]
  \Assert{t_1'} \Seq \Assert{t_2'} \Seq \Nondet_{(\Ata_1,c'_1) \in \C_1} \Nondet_{(\Ata_2,c'_2) \in \C_2} \\&
  (\Ata \together \Ata_1) \Seq ((\Ata^{j} \nondet \Ata^{j+1}) \Seq \Atb \Seq \Abort \together c'_1) \parallel 
  (\Ata \together \Ata_2) \Seq ((\Ata^{j} \nondet \Ata^{j+1}) \Seq \Atb \Seq \Abort \together c'_2)
 \Equals*[by atomic interchange \refax{sync-interchange-seq-atomic} for $\parallel$]
  \Assert{t_1'} \Seq \Assert{t_2'} \Seq \Nondet_{(\Ata_1,c'_1) \in \C_1} \Nondet_{(\Ata_2,c'_2) \in \C_2} \\&
  ((\Ata \together \Ata_1) \parallel (\Ata \together \Ata_2)) \Seq 
  (((\Ata^{j} \nondet \Ata^{j+1}) \Seq \Atb \Seq \Abort \together c'_1) \parallel 
   ((\Ata^{j} \nondet \Ata^{j+1}) \Seq \Atb \Seq \Abort \together c'_2))
 \Refsto*[by \reflem*{par-distrib-atomic} as $\Ata = \Ata \parallel \Ata$; inductive assumption]
  \Assert{t_1'} \Seq \Assert{t_2'} \Seq \Nondet_{(\Ata_1,c'_1) \in \C_1} \Nondet_{(\Ata_2,c'_2) \in \C_2} 
  (\Ata \together (\Ata_1 \parallel \Ata_2)) \Seq (\Ata^{j} \Seq \Atb \Seq \Abort \together (c'_1 \parallel c'_2))
 \Equals*[by atomic interchange \refax{sync-interchange-seq-atomic} for $\together$]
  \Assert{t_1'} \Seq \Assert{t_2'} \Seq \Nondet_{(\Ata_1,c'_1) \in \C_1} \Nondet_{(\Ata_2,c'_2) \in \C_2} 
  (\Ata \Seq \Ata^{j} \Seq \Atb \Seq \Abort \together (\Ata_1 \parallel \Ata_2) \Seq (c'_1 \parallel c'_2))
 \Equals*[as $\Ata \Seq \Ata^j = \Ata^{j+1}$ and \refax{sync-interchange-seq-atomic} for $\parallel$]
  \Assert{t_1'} \Seq \Assert{t_2'} \Seq \Nondet_{(\Ata_1,c'_1) \in \C_1} \Nondet_{(\Ata_2,c'_2) \in \C_2} 
  (\Ata^{j+1} \Seq \Atb \Seq \Abort \together (\Ata_1 \Seq c'_1 \parallel \Ata_2 \Seq c'_2))
 \Equals*[distribute by \reflem{sync-distrib-Nondet} for $\together$ and $\parallel$ and \refprop{assert-command-sync-command}] 
  \Ata^{j+1} \Seq \Atb \Seq \Abort \together 
  (\Assert{t_1'} \Seq \Nondet_{(\Ata_1,c'_1) \in \C_1} (\Ata_1 \Seq c'_1) \parallel \Assert{t_2'} \Seq  \Nondet_{(\Ata_2,c'_2) \in \C_2} (\Ata_2 \Seq c'_2))
 \Equals*[by \reflem{unrolled-atomic} and unrolled forms for $c_1$ and $c_2$]
  \Ata^{j+1} \Seq \Atb \Seq \Abort \together (c_1 \parallel c_2)
 \qedhere
\end{align*}
\end{proof}

\end{document}